# Ferrofluid bend channel flows for multi-parameter tunable heat transfer enhancement – Part 1: Numerical Modeling & Characterization

Authors: Nadish Anand[a,1] & Warren Jasper[a]

## Abstract

This study investigates ferrohydrodynamic heat transfer enhancement in a two-dimensional 90° bend channel through systematic parametric analysis of externally applied non-uniform magnetic fields, using Numerical CFD simulations. Two current-carrying wires positioned near the bend generate spatially varying magnetic fields that drive Kelvin body forces in the ferrofluid, creating localized mixing and boundary layer disruption. A comprehensive numerical analysis examines the configurations systematically varying six key parameters: Reynolds number (21 values, from 5 – 25), bend outer radius (3 values: 0.02, 0.04, 0.06 m), wire angle relative to horizontal (7 values from 30° to 60° in 5° increments), wire distance from bend center (2 values: 0.0075, 0.0100 m), nanoparticle volume fraction (2 values: 5%, 10%), and electromagnetic configuration of current carrying wires (9 current cases including single-wire, dual-wire aligned, dual-wire opposing, and baseline, of which 8 are symmetric to each other, effectively reducing this to 4 cases with current and 1 baseline case without current). Heat transfer performance is quantified using Average Nusselt numbers in four regions: whole channel, whole bend, first bend section, and second bend section. Results reveal that proper parameter selection achieves local heat transfer enhancements in the bend up to 300 to 400% compared to baseline flow. Reynolds number effects show optimal performance at Re = 5 (i.e. at the lowest Reynolds Number), with performance degrading 35% to 58% at Re = 20 as inertial forces overwhelm magnetic body forces. Bend radius exhibits strong coupling with other parameters, showing 10% to 44% performance reduction when curvature changes from tight ($R_o$ = 0.02 m) to gentle ($R_o$ = 0.06 m). Wire angle optimization reveals complex V-shaped behavior for opposing current configurations, with peak performance at angle extremes (30° or 60°) and 20% to 28% performance valleys at intermediate angles. Wire distance proves most critical, demonstrating universal negative dependence with 2% to 43% losses as distance increases 33%, following inverse square field decay. Nanoparticle concentration shows universal positive effects, with doubling from 5% to 10% producing 2% to 64% enhancements primarily through doubled magnetic susceptibility rather than modest thermal conductivity gains. The second bend section emerges as most responsive to nanoparticle concentration (7% to 64% range), while the first bend section achieves highest absolute performance levels. The optimal configuration employs opposing wire currents at 30° angle, tight bend radius, close wire positioning (0.0075 m), moderate Reynolds number (Re = 5), and high nanoparticle loading (10%), achieving Nusselt numbers exceeding 16 (almost 4x of the baseline Nusselt number) in the first bend section. The same configuration achieved highest overall Nusselt numbers in all the four

---

a – North Carolina State University
1 – Corresponding Author: nanand@alumni.ncsu.edu

examined regions, with global enhancement reaching 30% above baseline, and local to 400% over baseline. These findings establish that while optimal configurations follow established trends, there are a lot of differential variations in heat transfer enhancement, which are non-monotonic in nature, due to complex interplay between parameters and different flow regimes. These observations can be leveraged for targeted flow control/heat transfer enhancement applications.

## 1. Introduction

Ferrofluids are stable colloidal suspensions of nanoscale ferromagnetic particles dispersed in a carrier liquid such as water, oil, or other suitable fluids. These particles are typically coated with surfactants to prevent agglomeration and ensure uniform dispersion throughout the fluid volume[1].A distinctive feature of ferrofluids is their ability to respond to externally applied magnetic fields, enabling manipulation of their flow and thermal characteristics by varying the strength, geometry, or spatial gradients of these fields [2]. This tunable behavior has led to their adoption or investigation in diverse applications including seals, bearings, targeted cooling of electronic devices, biomedical systems, and advanced manufacturing processes [3].

Ferrofluids are used in a wide variety of systems for heat transfer, mixing, thermodynamic, energy harvesting, magnetorheological, etc.[4–8] applications. Heat transfer enhancement with externally controllable fluids is especially important in electronics cooling[9]. Various studies have been performed on ferrofluid/nanofluid convection in enclosures with different types of heat sources. However, recently, there has been much interest in self-pumping mechanisms and cooling through temperature-sensitive ferrofluids.[10–13].

The theoretical basis for ferrofluid dynamics termed Ferrohydrodynamics was pioneered by Rosensweig and Shliomis, who formulated models describing the coupling between magnetic field forces and fluid motion [1,2,14,15]. Under a non-uniform magnetic field, the magnetized particles experience a body force, referred to as the Kelvin body force, which drives the fluid toward regions of higher magnetic field intensity [3,16]. The magnitude of this force depends on the field gradient, particle concentration, particle magnetic properties, and the carrier fluid's viscosity and density [17–20]. Such forces can be induced using permanent magnets or electromagnets, which allow for more dynamic control using varying current.

Experimental and numerical studies demonstrate significant heat transfer enhancement in channels under magnetic field control. Asfer et al. [21] conducted infrared thermography experiments on ferrofluid flow through circular stainless-steel tubes with 2-millimeter diameter at low Reynolds numbers, showing that convective heat transfer coefficients can increase or decrease depending on magnetic field configuration, particle aggregation effects, and the ratio of magnetic force to inertia force. Their COMSOL simulations captured particle migration and chain formation along field lines. Aminfar et al. [22] performed numerical investigations of curved tube geometries, revealing complex interactions between curvature-induced centrifugal forces known as Dean vortices and magnetic body forces, with optimal enhancement occurring when magnetic field gradients align with thermal gradients.

Ganguly et al. [23] investigated thermomagnetic convection in square enclosures using line-dipole magnetic fields, demonstrating heat transfer augmentation through localized recirculation zones induced by Kelvin forces. The study established scaling relationships between magnetic Rayleigh number and heat transfer enhancement, providing design guidelines for practical magnetoconvection systems. Mukhopadhyay et al. [24,25] developed comprehensive scaling analysis characterizing thermomagnetic convection, introducing dimensionless groups that govern convection onset and demonstrating that thermomagnetic effects dominate when the magnetic Rayleigh number exceeds the thermal Rayleigh number.

Curved channel geometries introduce centrifugal effects that synergistically interact with magnetic forcing. Sheikholeslami et al. [26] performed finite volume simulations of ferrofluid flow in 90° elbow channels with non-uniform magnetic fields from current-carrying wires. Their SIMPLEC algorithm results demonstrated average Nusselt number enhancement of 28.6% at magnetic number of $9.32 \times 10^6$ and Reynolds number of 50, with local enhancements exceeding 200 percent near magnetic source at approximately 45° of the bend. The magnetic field induced secondary flow circulations near the inner elbow wall, with circulation intensity decreasing at higher Reynolds numbers as inflow momentum overcame magnetic forcing. Heat transfer also increased 18% when Reynolds number rose from 50 to 100 due to enhanced convective transport. Colombo et al. [27] analyzed fully developed laminar mixed convection in horizontal annular ducts using Galerkin spectral methods, providing benchmark solutions for complex curved geometries that serve as validation cases for ferrofluid simulations.

Complex geometries with multiple magnetic sources enable sophisticated thermal control. Gerdroodbary et al. [28] investigated T-junction flows with three current-carrying wires, achieving average heat transfer enhancement exceeding 64% with local peaks surpassing 200 % near the first magnetic source. Sinusoidal double-pipe configurations [29] combine passive geometric enhancement through wavy walls with active magnetic control, showing that amplitude coefficient and magnetic field intensity act synergistically. Nusselt numbers increased 25% with amplitude changes and an additional 23 to 50% under magnetic fields. Twisted tape inserts provide further augmentation: Mokhtari et al. [30] examined swirling ferrofluid flow in tubes with parallel current-carrying wires, finding twisted tapes alone increased Nusselt numbers by 200% while magnetic fields added 30% enhancement. The combination creates spiral flow patterns with enhanced turbulent mixing and magnetically induced secondary flows. Goharkhah and Ashjaee [31] studied alternating magnetic fields, observing friction factor increases of 1.2 to 1.8 times depending on field strength and frequency, with maximum heat transfer enhancement of 13.9% at Reynolds number of 2000 and frequency of 20 Hertz.

Microscale applications leverage thermomagnetic convection for passive cooling without mechanical pumps. Chaudhary et al. [32] developed self-pumping magnetic cooling devices using manganese-zinc ferrite nanoparticles at 10% concentration synthesized via hydrothermal methods, achieving temperature drops of 20 to 28 °C under 0.3 Tesla magnetic fields from neodymium-iron-boron permanent magnets with zero external pumping power. The system exhibited self-regulating behavior: as heat loads increase, ferrofluid magnetization decreases due to the pyromagnetic effect where magnetization decreases with increasing temperature, inducing stronger Kelvin force gradients and faster circulation for enhanced heat removal. Combined experimental measurements and COMSOL simulations validated the thermomagnetic convection mechanism. Zablotsky et al. [33] investigated surface cooling based on thermomagnetic convection through combined numerical simulation and experiments, demonstrating maximal intensification when magnets are positioned near warm ends and achieving cooling effects of 0.075 W/cm$^2$.

Transformer cooling and electronics thermal management represent major application areas. Zanella et al. [34] performed comprehensive spectral-finite element simulations using the Magnetohydrodynamics code called SFEMaNS for ferrofluid-cooled cylindrical transformer geometries with immersed coils, validating against experimental data with vegetable-based transformer oil. Their axisymmetric simulations examined thermomagnetic convection via Kelvin force effects and altered thermophysical properties from iron oxide nanoparticle addition at volume fractions from 0 to 7%. Maximum temperatures decreased from 71.2 °C with pure oil to 67.1°C with 7% ferrofluid without ferromagnetic cores. When cores with relative permeability of 1000 were added, temperature drops reached 12°C compared to 2°C without cores, as high permeability cores amplify magnetic field intensity from approximately 15 kA/m to approximately 45 kA/m in the fluid region. Banerjee et al. [35] conducted numerical optimization for electronics cooling in square cavities with discrete heaters and line-dipole magnetic fields, determining safe operating temperature ranges by varying heater configurations and magnetic field positions.

Computational modeling requires sophisticated coupling of fluid dynamics, thermal transport, and electromagnetics. Strek [36] analyzed ferrofluid channel flow using COMSOL-based finite element methods, demonstrating that strategically positioned magnetic dipoles induce magnetoconvection when Kelvin forces overcome viscous damping. Temperature-sensitive magnetic fluids in square cavities show distinct convection patterns under uniform magnetic fields [37], with heat transfer enhancement depending on Rayleigh number, magnetic coefficient, and nanoparticle volume fraction. Numerical simulations must account for temperature-dependent properties including viscosity, thermal conductivity, and magnetization, and validate against experimental measurements to capture nonlinear coupling effects accurately. Lajvardi et al. [38] experimentally investigated enhanced ferrofluid heat transfer under magnetic field effects, finding significant improvements in thermal conductivity and convective performance with increasing field strength.

Advanced geometries and flow configurations extend ferrofluid thermal management capabilities. Szabo and Früh [39] studied the transition from natural convection to thermomagnetic convection in non-uniform magnetic fields, identifying critical parameters for onset of magnetic forcing dominance through numerical simulations validated against experimental data. Schumacher et al. [40] investigated turbulent ferrofluid channel flow with magnetic fields and particle rotation effects using direct numerical simulation, demonstrating pressure drop penalties accompanying heat transfer enhancement. Their direct numerical simulations captured Reynolds stress distributions and vorticity dynamics under magnetic influence, revealing complex interactions between turbulent fluctuations and magnetic body forces that modify flow structure and thermal transport. Hangi and Bahiraei [41] employed two-phase Eulerian-Lagrangian simulations for ferrofluid flow between parallel plates under localized magnetic fields, tracking individual particle trajectories and demonstrating concentration non-uniformity and particle migration toward high-field regions.

Nanoparticle characteristics profoundly influence performance. Ortiz-Salazar et al. [42] investigated magnetic field-induced tunability of thermal conductivity in ferrofluids loaded with carbon nanofibers, demonstrating that hybrid nanostructures combining magnetic nanoparticles with carbon nanomaterials create synergistic effects. Magnetic fields align nanofibers into high-conductivity pathways while magnetic nanoparticles respond to thermomagnetic forcing. Advanced corrugated and wavy channel designs combine geometric and magnetic enhancement mechanisms. Asadi et al. [43] numerically investigated ferrofluid flow in wavy channels under non-uniform magnetic fields, finding that wave amplitude and negative magnetic field gradients synergistically enhance Nusselt numbers through induced vortices and secondary flows. Larger wave amplitudes with stronger negative gradients maximize heat transfer but incur pumping power penalties. Many studies [44–49] on magnetic nanofluid applications in thermal engineering demonstrate continued innovation in passive magnetic enhancement techniques for heat exchangers, thermal management systems, and cooling applications across diverse Reynolds number regimes, magnetic elements, and geometric configurations.

This work presents a numerical analysis of heat transfer enhancement in a 90° elbow channel. The analysis is performed on three different 90° elbows with outer radii of the bend equal to two, four and six channel widths. The ferrofluid flow inside the channel is controlled by two parametrically placed wires at a parametric distance from the center of the bend. These wires can carry current either into or out of the plane and have different magnitudes. The non-uniform magnetic force generated by the wires will exert a Kelvin body force on the ferrofluid.

Furthermore, the system is configured to allow for cooling system design with an additional angular variable, which can be adjusted to change the cooling needs for the channel. This angular variable helps maintain the distance from the center point of the channel while providing an extra degree of freedom to alter the magnetic field distribution inside the channel. The magnetic field alteration will be responsible for different flow behavior and thus may increase/decrease local or global cooling performance. The cooling performance is characterized by a set of four Nusselt numbers, which characterize the system's local and global cooling performance. In the next section, the configuration of the system is described; in section 3, the governing equations along with the constitutive relations and thermophysical properties are discussed; section 4 provides the details of numerical modeling; section 5 presents the results of the numerical modeling, & section 6 summarizes the findings and draws conclusions.

## 2. System Configuration

**Figure 1** and **Figure 2** represent the system configuration for the elbow channel with outer radius of 2 channel widths from the bend center point. The outer domain is kept sufficiently large as shown in **Figure 1**, which will allow for the convergence of magnetic field.

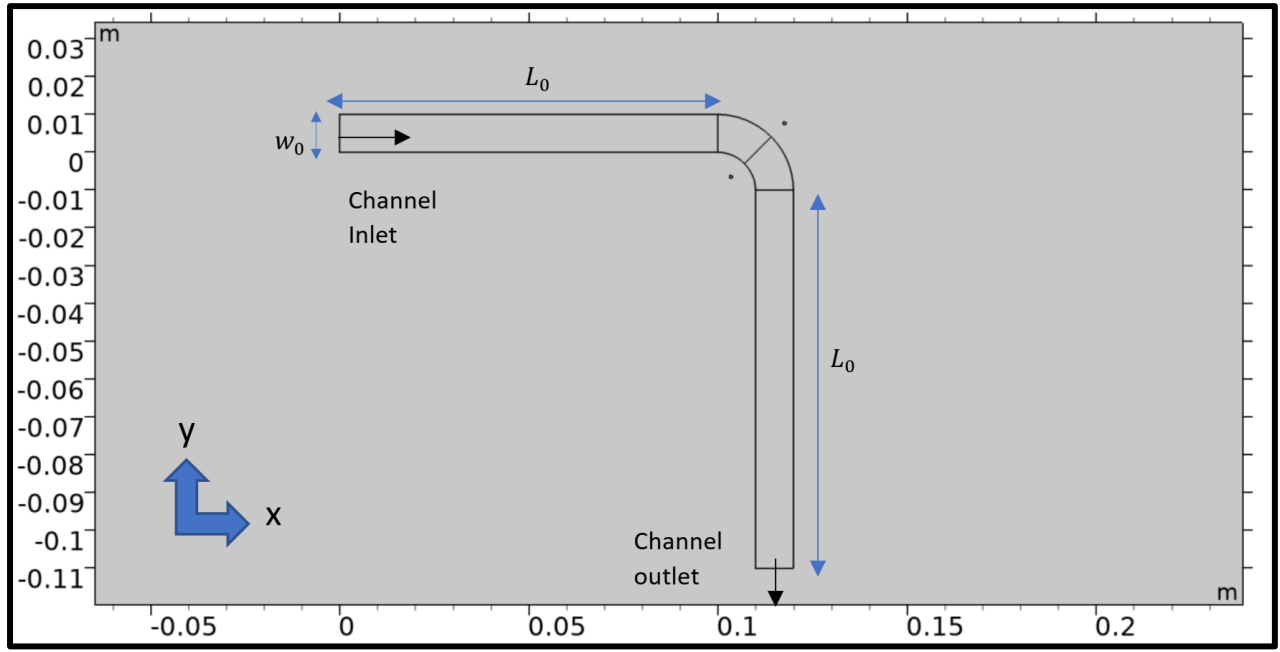

Figure 1 Configuration of the elbow channel

The lengths of the two channels before and after the bend is $L_0$, which is set to 10 channel widths, to give enough space for the flow to stabilize. The wires can carry current in both directions i.e., into and out of the plane. These cases are discussed in detail in section 4.

In **Figure 2** the parametric placement of the wires is shown along with the bend inlet, outlet, and the bend center. The bend center is chosen such that it divides the bend and by extension the channel into two parts of equal area, separated by an angle of 45°.

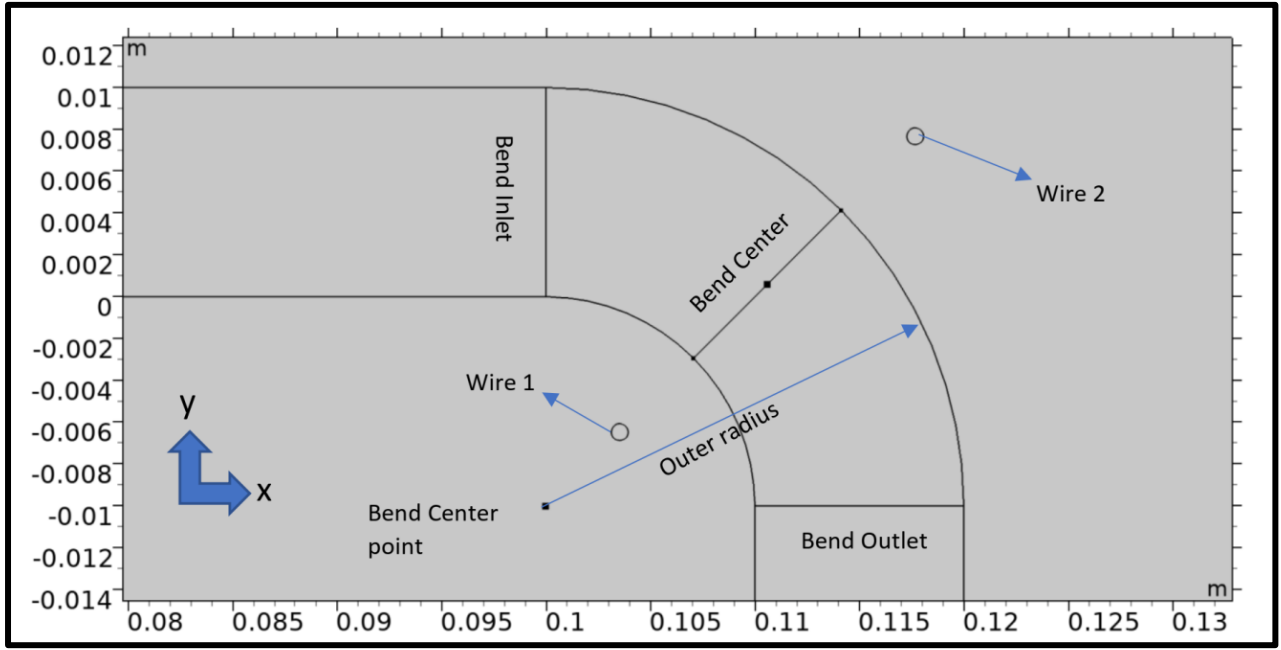

**Figure 2** Elbow channel bend for outer radius of 2 channel widths

In this study three channels are studied. All elbow channels undergo a 90° bend, however, differ from each other by the magnitude of the outer radius. The first channel has an outer radius of $2w_0$ while the second has an outer radius of $4w_0$ and the third channel has an outer radius of $6w_0$. Another variation is within the angle which the line joining the two wires make with the horizontal. **Figure 3** shows three different cases of symmetrical angular displacement of the wires with respect to the first two channel centers. It must be noted thar how the channel curvature becomes smooth by increasing the outer radius ($R_o$).

The wires are always kept at a distance of $dist_1$ from the center of the bend and are given an angular displacement of $\alpha = 30°\ to\ 60°, in\ 5°\ increments$. The radial displacement of the wires from the bend center is called as $dist_1$ and is also parametrized, details of which are discussed in Section 4.

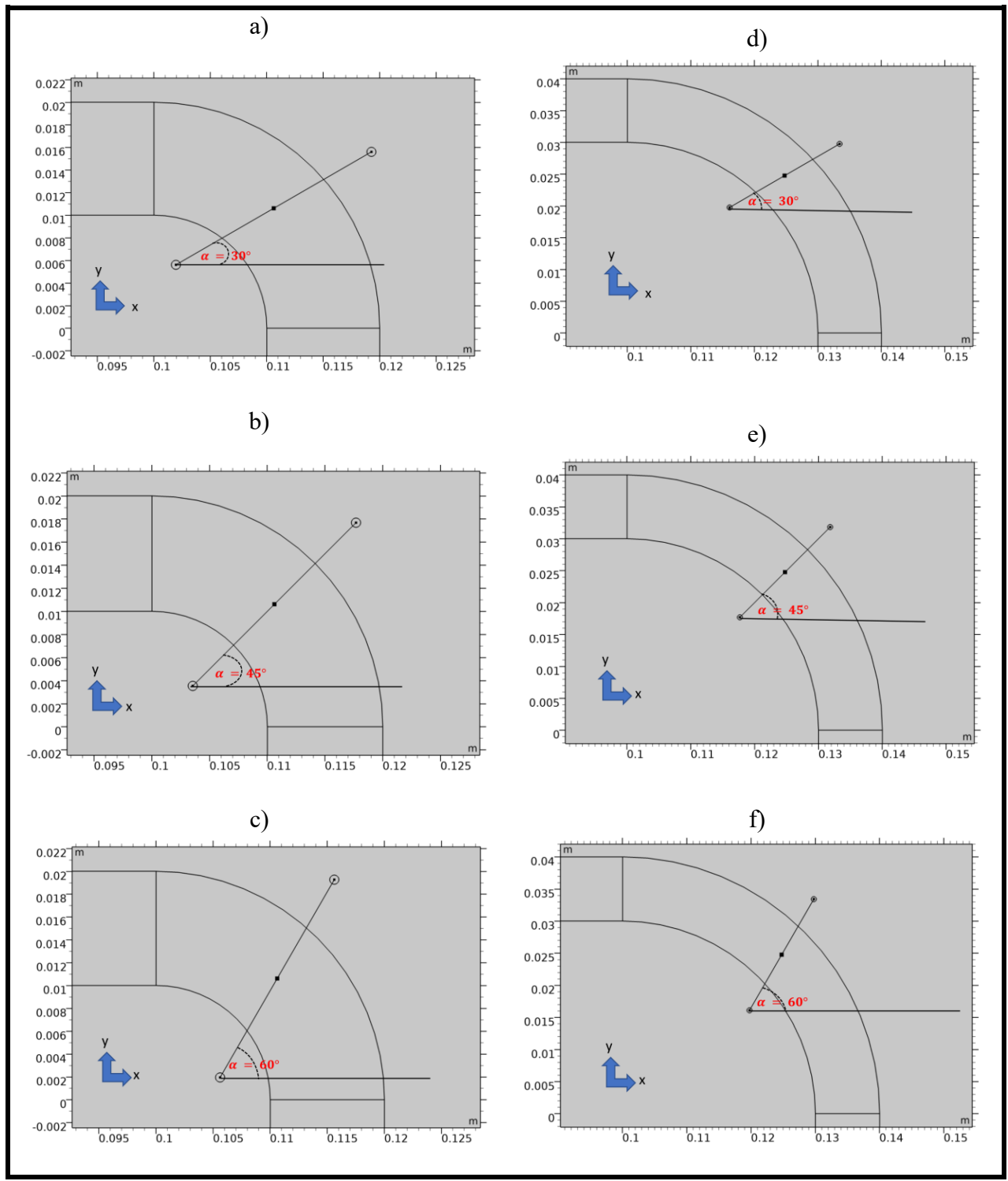

**Figure 3** Angular position of the wires with respect to the horizontal for a),b) and c) outer radius of $2w_0$ and d),e) and f) outer radius of $4w_0$

# 3. Governing Equations

The general continuity, momentum and energy equations govern the ferrofluid flow through the channels. The following assumptions are made for the fluid and the flow:

1. The ferrofluid is linear, i.e. the magnetization is linearly related to the magnetic field strength
2. Ferrofluid is a Newtonian fluid.
3. Ferrofluid is non-conductive.
4. Steady state, laminar, irrotational and incompressible flow.
5. No internal heat generation.
6. The ferrofluid is not temperature sensitive (Magnetocaloric effect is not considered).

The ferrofluid is assumed to be a colloidal solution of 11 nm diameter Magnetite ($Fe_3O_4$) nanoparticles and de-ionized water. The concentration of ferrofluid is varied from 5% to 10% and since the Curie temperature of magnetite is around 850 K [50,51], it is a good assumption to neglect the temperature dependence of physical properties.

The continuity and momentum equations are given by (1)-(3) as follows:

Continuity:

$$\frac{\partial u}{\partial x} + \frac{\partial v}{\partial x} = 0 \qquad (1)$$

x and y Momentum:

$$u\frac{\partial u}{\partial x} + v\frac{\partial u}{\partial y} = -\frac{1}{\rho}\frac{\partial p}{\partial x} + \nu\left(\frac{\partial^2 u}{\partial x^2} + \frac{\partial^2 u}{\partial y^2}\right) + \frac{\mu_0}{\rho}\left(M_x\frac{\partial H}{\partial x} + M_x\frac{\partial H}{\partial y}\right) \qquad (2)$$

$$u\frac{\partial v}{\partial x} + v\frac{\partial v}{\partial y} = -\frac{1}{\rho}\frac{\partial p}{\partial y} + \nu\left(\frac{\partial^2 v}{\partial x^2} + \frac{\partial^2 v}{\partial y^2}\right) + \frac{\mu_0}{\rho}\left(M_y\frac{\partial H}{\partial y} + M_y\frac{\partial H}{\partial x}\right) \qquad (3)$$

The third term on the RHS in equations (2) and (3) are the Kelvin body force exerted on the ferrofluid due to non-uniform magnetic field. The ferrofluid magnetization and magnetic fields are assumed to be sensibly collinear, and hence there is no additional viscous term as was observed in the previous chapter, where a set inlet magnetization was imparted.

Energy Equation with a constant heat flux is given as:

$$u\frac{\partial T}{\partial x} + v\frac{\partial T}{\partial y} = \frac{1}{\rho C_p}\left(\frac{\partial^2 T}{\partial x^2} + \frac{\partial^2 T}{\partial y^2}\right) + \frac{\eta}{\rho C_p}\left\{2\left[\left(\frac{\partial v}{\partial y}\right)^2 + \left(\frac{\partial u}{\partial x}\right)^2\right] + \left(\frac{\partial v}{\partial x} + \frac{\partial u}{\partial y}\right)^2\right\} \qquad (4)$$

The following constitutive relations hold for the magnetic field:

$$\nabla \cdot \vec{B} = 0 \qquad (5)$$

$$\nabla \times \vec{H} = 0 \tag{6}$$

$$\vec{M} = \chi_m \vec{H} \tag{7}$$

Since the ferrofluid is assumed to be generic, in order to find the material properties of the dissolved phase of the ferrofluid the following property relations, in terms of the volume fraction, $\phi$, are used for the thermal conductivity, heat capacity, density and magnetic susceptibility of the ferrofluid [52–57]:

Thermal conductivity:

$$k = \frac{\left(2k_L + k_p - 2\phi(k_L - k_p)\right)}{k_L + k_p + \phi(k_L - k_p)} k_L \tag{8}$$

Density:

$$\rho = (1 - \phi)\rho_L + \phi\rho_p \tag{9}$$

Heat capacity:

$$\rho C_p = (1 - \phi)(\rho C_p)_L + \phi(\rho C_p)_p \tag{10}$$

Viscosity:

$$\eta = \frac{\eta_L}{(1 - \phi)^{2.5}} \tag{11}$$

Magnetic susceptibility:

The initial Langevin susceptibility is given by:

$$\chi_i = \frac{\mu_0 \phi M_d m}{3k_B T_{ref}} \tag{12}$$

The magnetic moment of one magnetite molecule is 4 Bohr Magneton, which can be converted into volumetric quantity through[58,59]:

$$m = \frac{4\mu_B \pi d^3}{6 \times 91.25 \times 10^{-30}} \tag{13}$$

The modified Langevin susceptibility for the ferrofluid is given by[53]:

$$\chi = \chi_i \left(1 + \frac{\chi_i}{3}\right) \tag{14}$$

The boundary conditions applied are:

Fluid flow is considered to be hydrodynamically fully developed as it enters the channel. The flow enters the channel with a preset Reynolds number, which determines the inlet velocity. On the channel walls, the flow has a 'no-slip' boundary condition. For magnetic field, since the magnetic susceptibility sharply varies at the air and channel interface (i.e. channel wall), the continuity of the normal B field and tangential H field is accounted for. To this effect, the walls of the channel are considered perfect magnetic conductors. This, combined with the constitutive relations described above, maintains the continuity of the field. Moreover, the outer domain shown in Figure 4 has the magnetic insulation boundary condition.

Finally, for the thermal boundary conditions, the whole of the channel walls is subjected to a net inward heat flux ($q''$) of 1000 W/m², with the inlet temperature set at 300 K. The reference temperature is set at 300 K.

In order to quantify the heat transfer a Nusselt number is defined as:

$$Nu = q'' \times \frac{w_0}{(T_{wall} - T_{mean}) \times k} \tag{15}$$

Where, the mean temperature $T_{mean}$ is defined as:

$$T_{mean} = \frac{\iint T.U \, dA}{\iint U \, dA} \tag{16}$$

Where A is the surface area of the channel and U the velocity magnitude.

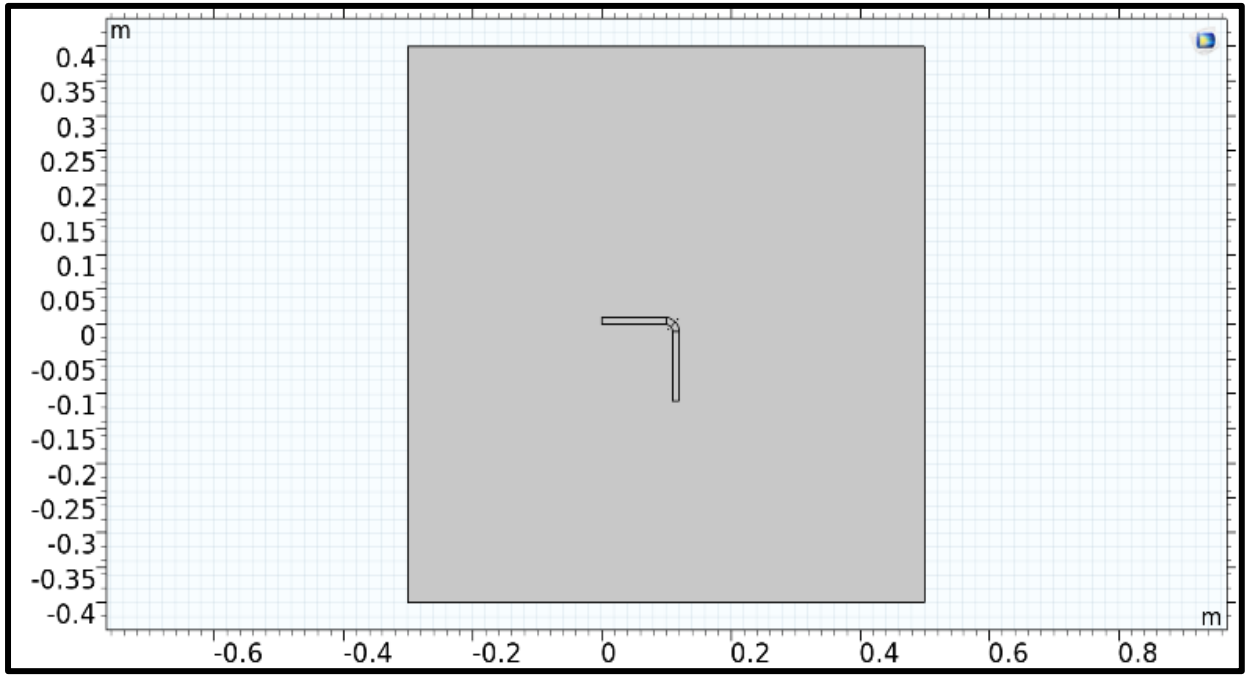

**Figure 4** Full Geometry with the extended domain for convergence of Magnetic Field

Finally, the Reynolds number is defined as:

$$Re = \frac{Uw_0}{\eta} \tag{17}$$

**Table 1** summarizes the properties of the ferrofluid for ϕ = 0.05 & 0.1.

Table 1 Ferrofluid physical properties

| Property | Magnetite Particles | Water | Ferrofluid ($\phi = 0.05$) | Ferrofluid ($\phi = 0.1$) |
|---|---|---|---|---|
| Viscosity [Pa. s] | none | 0.001 | 0.0011368 | 0.0013013 |
| Density [kg/m$^3$] | 5100 | 1000 | 1205 | 1410 |
| Thermal conductivity [W/m-K] | 7 | 0.585 | 0.7096 | 0.7964 |
| Specific Heat [J/kg-K] | 660 | 4180 | 3435.1 | 2885.11 |
| Magnetic Susceptibility | - | - | 0.72803 | 1.702 |

In order to study the effects of the non-uniform magnetic field generated by the parametric placement of the wires for both the channels different parameters are chosen. Then the system is numerically modeled and 4 Nusselt numbers are calculated. These Nusselt numbers are for 4 different parts of the channel. They are as follows:

a.) Nusselt number for the whole channel

b.) Nusselt number for the whole bend

c.) Nusselt number for the first bend section (i.e. from bend inlet to bend center)

d.) Nusselt number for the second bend section (i.e. from bend center to bend outlet)

The choice of these Nusselt numbers allows for the quantification of heat transfer enhancement at local and global levels.

## 4. Numerical Modeling

COMSOL Multiphysics version 5.4 is used for numerically simulating the set of governing equations discussed in the previous section[60]. COMSOL is Finite Element Method (FEM) based software which solves conservation laws using the weak form and weighting functions of specified order. COMSOL uses Newton's method to solve the partial differential equations and for Navier-Stokes equation and uses stabilization methods based on Streamline Upwind Petrov Galerkin (SUPG) methods, which resolves the issue of oscillation of solutions in cases where upwinding is not used[10,61–64]. The internal multiphysics interfaces/modules to couple magnetic and flow fields is utilized. Second order elements are used to discretize both pressure and velocity fields. The magnetic field is discretized using quadratic elements.

Since the inlet velocity is uniform entering the channel it can be expressed in terms of the Reynolds number. The solution procedure employed uses a segregated approach in which different dependent variable groups are solved simultaneously but separately within one iteration of the whole solution. The PARDISO algorithm is used to solve for all the segregated groups within the segregated iteration of the solution for all the internal groups.

The dependent variables solved for in groups are as follows:

a.) Group 1: Velocity and Pressure

b.) Group 2: Temperature

c.) Group 3: Magnetic vector potential, wire current and wire voltage.

A convergence criterion is set in tolerance to be $10^{-5}$ for the solution and residual for all the groups and the solution in one main iteration of the solution. This is in general considered to be a very tight convergence criterion[65,66]. The parameters which were varied are stated in **Table 2**.

Table 2 Parameter varied for simulation and their values/limits

| Parameter | Values/Limits |
|---|---|
| Current in wire 1 ($I_1$) | -5, 0 and 5 [A] |
| Current in wire 2 ($I_2$) | -5, 0 and 5 [A] |
| Distance between wires and bend center ($dist_1$) | $0.75w_0$ & $w_0$ |
| Outer Radius of the bend ($R_o$) | $2w_0$ (channel 1), $4w_0$ (channel 2) and $6w_0$ (channel 3) |
| Angle of wires with horizontal ($\alpha$)1 | 30° to 60°, in 5° advancements (7 values) |
| Reynolds Number ($Re$) | 5 to 25 |
| Ferrofluid concentration ($\phi$) | 5% and 10% |

Since the current in the two wires have three values each, there would be 9 different cases of wire current as listed in **Table 3Error! Reference source not found.**.

Table 3 Cases for current flowing in wire 1 and wire 2

| Case Number | $I_1$ | Direction for $I_1$ | $I_2$ | Direction for $I_2$ |
|---|---|---|---|---|
| 1 | -5[A] | Into the plane | -5[A] | Into the plane |
| 2 | -5[A] | Into the plane | 0[A] | none |
| 3 | -5[A] | Into the plane | 5[A] | Out of the plane |
| 4 | 0[A] | none | -5[A] | Into the plane |
| 5 | 0[A] | none | 0[A] | none |
| 6 | 0[A] | none | 5[A] | Out of the plane |
| 7 | 5[A] | Out of the plane | -5[A] | Into the plane |
| 8 | 5[A] | Out of the plane | 0[A] | none |
| 9 | 5[A] | Out of the plane | 5[A] | Out of the plane |

Grid convergence studies were performed on a modified grid with $N_{cell}$ going from 10 to 50, for all the three channels, for all the parameters and for the 9 current cases. It was observed that $N_{cell} = 40$ provides a very accurate solution with less than 2% variation of the Nusselt number values for the channel and bend compared to those calculated from $N_{cell} = 50$. Hence a value of $N_{cell} = 40$ is used in the simulations.

## 5. Results and Discussions

The heat transfer characteristics are analyzed using the 4 Nusselt numbers as described in the previous section. Each of the 9 cases are analyzed for each case of the angle $\alpha$ to obtain a comparison between the Nusselt numbers. **Figure 5** shows the velocity magnitude for all the 9 cases summarized in **Table 3** for Reynolds number 5 and outer radius of the channel $2w_0$ (which is referred to as channel 1), when $\alpha$ is set to 45°.

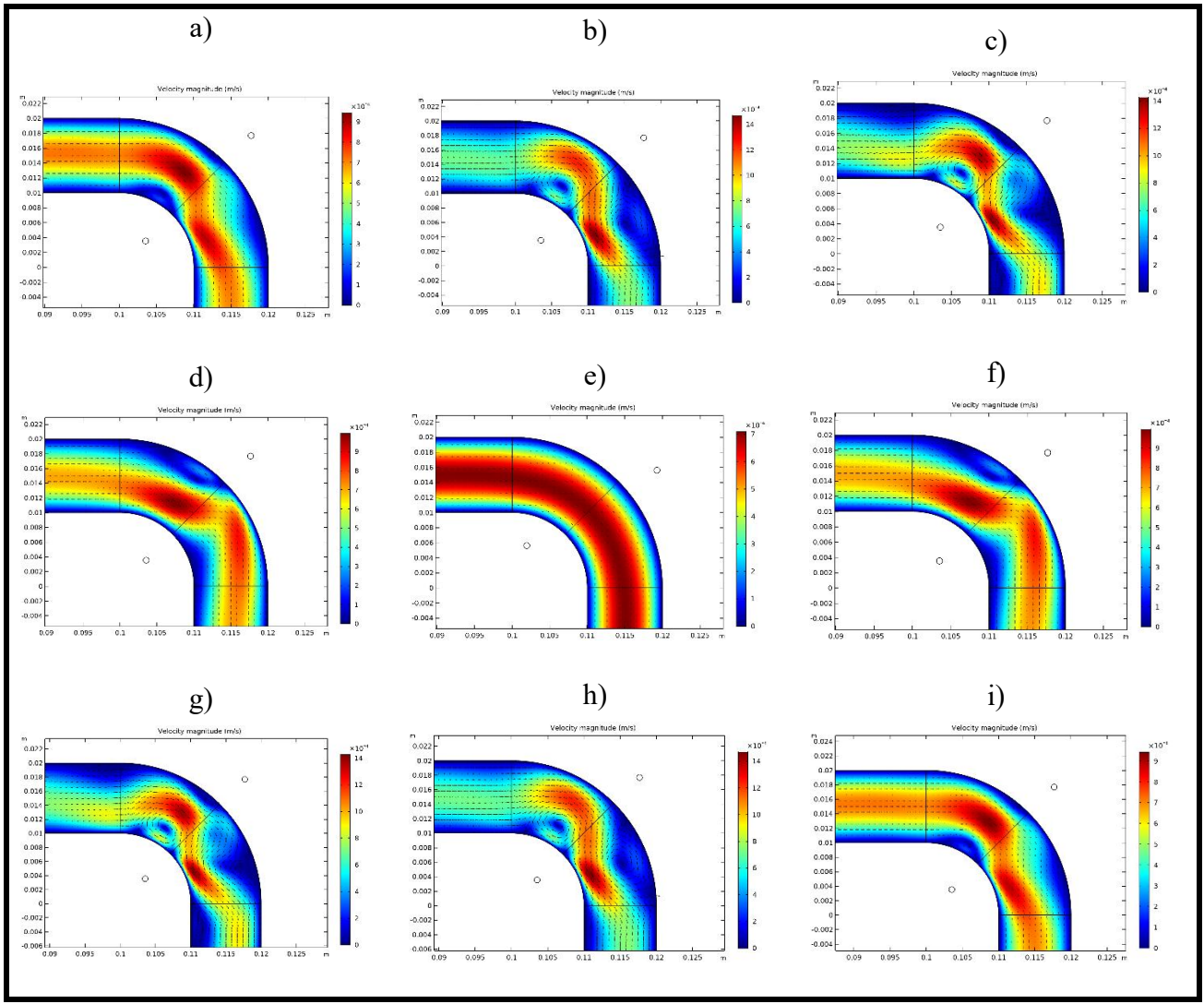

**Figure 5** Velocity magnitude plot for $2w_0$ channel for $\alpha = 45°$ for Re = 5 for a) Case 1 b) Case 2 c) Case 3 d) Case 4 e) Case 5 f) Case 6 g) Case 7 h) Case 8 i) Case 9

The fifth case shown in **Figure 5**, is where no current flows through the wires, and hence is the reference case. It is observed from the velocity magnitude plots that there is almost identical behavior exhibited by cases 1 & 9, cases 2 & 8, cases 3 & 7 and cases 4 & 6. Hence, the 9 cases are reduced ultimately into 5 including one reference case. The cases 1&9, 2&8, 3&7, and 4&6 overlap completely, due to symmetry in the Magnetic field[5]. Hence, all further analysis would be performed based on the first 5 cases mentioned in **Table 3**, including the reference case of no Kelvin Body force when the magnetic field is absent. In the first sub-section, the variation of all the 4 Nusselt numbers is discussed with Reynolds number and all the 5 current cases. In the second sub-section, variation of the 4 Nusselt numbers with Outer Radius is discussed, in the third sub-section, variation of the Nusselt numbers with wire angles is discussed, in the fourth sub-section the variation of the Nusselt Numbers with wire distance is discussed and finally in the fifth sub-section variation of the Nusselt numbers with the Ferromagnetic particle volume fraction is discussed. The analysis focuses on finding general trends and observations, with the breadth of variations captured in the sub-sections.

## 5.1 Effect of Reynolds number on Heat transfer enhancement

The influence of the current carrying wire configuration on ferrofluid heat transfer characteristics was systematically investigated across varying Reynolds numbers (Re = 5–20) and other parameters. Reynolds number controls the balance between inertial and viscous forces in the flow, directly affecting how magnetic body forces interact with fluid motion. All analyzed configurations in this section use tight bend curvature ($R_o = 0.02\ m$), close wire positioning ($dist_1 = 0.0075\ m$), and 5% nanoparticle concentration. Three wire angles, $\alpha$ (30°, 45°, 60°) demonstrate how field orientation affects the Reynolds number response. At low Reynolds numbers, viscous forces dominate and magnetic body forces can effectively modify the flow field. As Reynolds number increases, inertial forces become stronger and the relative influence of ferrohydrodynamic mixing diminishes. **Figure 6** shows heat transfer behavior at the shallowest wire angle of 30° with close wire positioning. This configuration creates strong magnetic field components aligned with the primary flow direction, which is optimal for ferrohydrodynamic mixing enhancement at low flow rates. The whole channel displays relatively flat Reynolds number dependence across most cases, with variations remaining within 10% to 15% across the entire Reynolds number range. Case 3 with opposing wire currents maintains the highest performance throughout, achieving Nu around 4.75 at Re = 5 and declining gradually to 4.52 at Re = 20, representing approximately 5% reduction. Cases 1, 2, and 4 cluster closely together, all maintaining Nu values between 4.3 and 4.55 across the Reynolds number range with modest variations. The baseline Case 5 shows slight positive trend from 4.17 to 4.33, indicating that purely hydrodynamic effects in the channel improve modestly with increasing Reynolds number, due to increased channel length needed for the flow to be thermally fully developed. It must also be noted that, in the formulation for the Nusselt number as state in equation (15), the Nusselt number is inversely proportional to temperature difference. Hence, the Nusselt number must increase as the temperature difference decreases, which happens for higher flow rates, and by extension for higher

Reynold's numbers. The bend region exhibits strong negative Reynolds number dependence across all ferrohydrodynamic cases, demonstrating that curved section performance is critically sensitive to flow regime. Case 3 dominates at low Reynolds numbers, achieving Nu around 9.5 at Re = 5 (which is around 220% increase above baseline Case 5), but experiences substantial decline to approximately 5.2 at Re = 20, representing dramatic 47% performance loss. This severe degradation occurs because centrifugal forces in the bend increase with the square of velocity, and at high Reynolds numbers these inertial effects overwhelm the magnetic body forces. Cases 1, 2, and 4 show similar steep negative trends, with Case 1 declining from 6.5 to 5 (25% loss), Case 2 from 6.5 to 4.8 (25% loss), and Case 4 from 7.0 to 5.0 (30% loss).

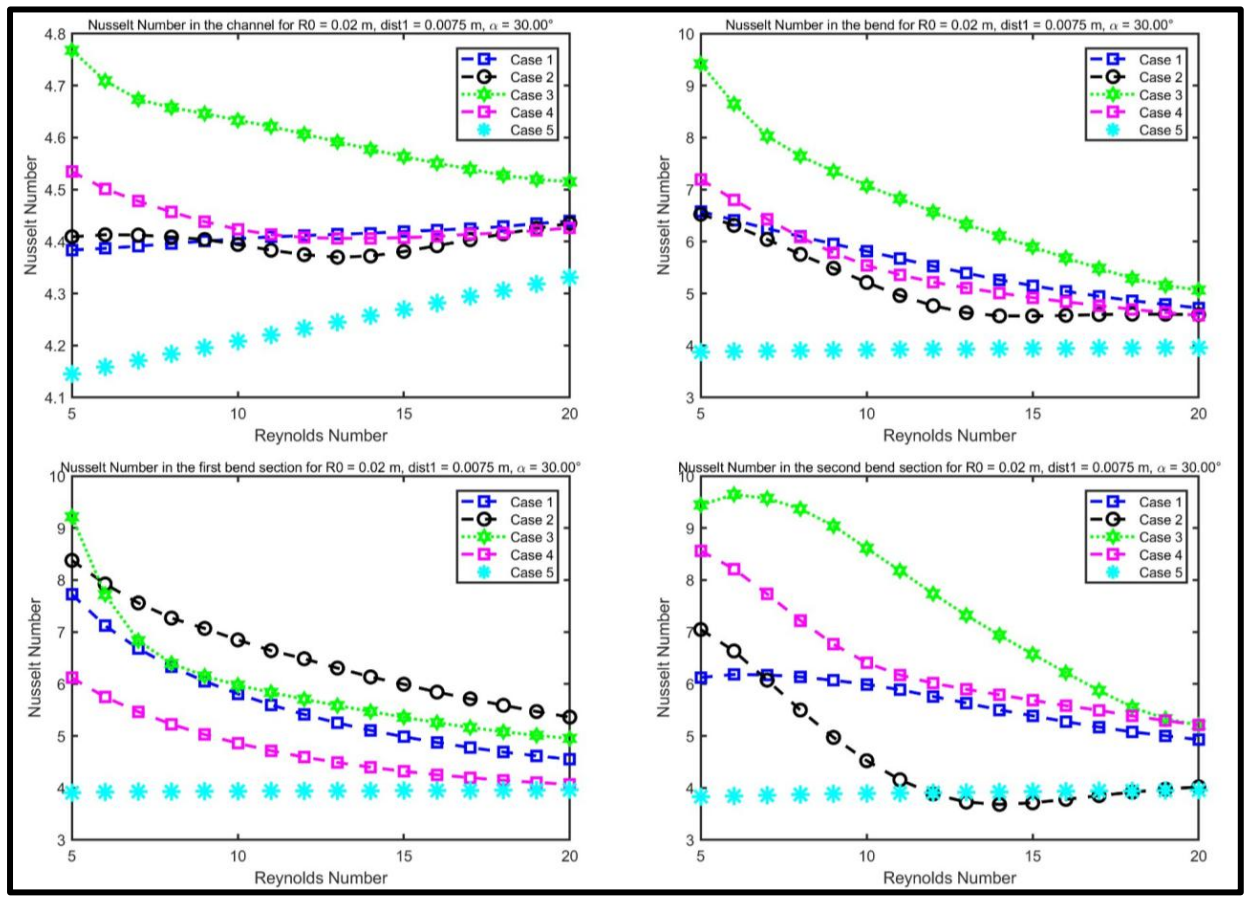

**Figure 6** Nusselt number variation with Reynolds number for $R_o = 0.02\ m$, $dist_1 = 0.0075\ m$, $\alpha = 30°$, $\phi = 0.05$.

All ferrohydrodynamic cases converge toward Nu values around 4.8 to 5.2 at Re = 20, indicating that high inertia conditions substantially reduce magnetic field enhancement and minimize case differentiation. The first bend section reveals steep Reynolds number sensitivity. Case 2 exhibits exceptional performance at low Reynolds numbers, achieving Nu around 8.5 at Re = 5, but experiences substantial decline to 5.5 at Re = 20, representing 35% performance loss. This entry zone where boundary layers are developing proves vulnerable to increasing inertial forces. Case 3 shows similar behavior, declining from 9.2 to 4.9 (47% loss), while Case 1 drops from 7.7 to 4.7 (39% loss) and Case 4 from 6.0 to 4.2 (30% loss), converging with Case 5, essentially overcoming all Ferrohydrodynamic mixing. The substantial Reynolds sensitivity in the first bend section establishes this as the region where ferrohydrodynamic enhancement is critically dependent on maintaining low to moderate flow rates. The second bend section demonstrates continued strong Reynolds number dependence. Case 3 maintains exceptional performance at low Reynolds numbers, achieving Nu around 9.4 at Re = 5, which increases to 9.7 at Re = 6. However, it experiences substantial 48% decline to 5.0 at Re = 20. Case 4 shows decline from 8.5 to 5.3 (38% loss), Case 1 from 6.2 to 5.0 (19% loss), while Case 2 shows moderate decline from 7.0 to 3.8 (46% loss). These wake region patterns establish that recirculation driven heat transfer enhancement, while substantial at low Reynolds numbers, diminishes progressively as flow inertia increases. At intermediate wire angle of 45° (**Figure 7**), Reynolds number effects show distinctly different characteristics compared to shallow angles, revealing how balanced field components alter flow regime sensitivity. The whole channel displays interesting behavior where different cases respond differently to Reynolds number increase. Case 2 & Case 4 shows negative trend until intermediate Reynolds Numbers and then further catching on with the general trend of the no-current case. While Case 1 shows modest positive trend rising from 4.28 to 4.37 (2% modest increase), Case 3 exhibits similar non-monotonic behavior to Cases 2 & 4 but troughing at Re = 8, before following the uptrend ending up highest in all the cases at Re = 20. By Re = 20, all cases converge closely to Nu values between 4.36 and 4.45, although non-monotonically, indicating that field orientation differences matter progressively less as inertial effects dominate. The baseline increases from 4.16 to 4.32, showing that all the Ferrohydrodynamic cases, with non-zero currents remain above the baseline values. The bend region demonstrates strong Reynolds number sensitivity with all cases experiencing performance degradation. Case 2 & Case 4 lead at low Reynolds numbers with Nu around 7 at Re = 5 but suffers 35% decline to 4.2 ~ 4.5 at Re = 20. Case 3 shows declines from approximately 6.3 at Re = 5 to 4.7 at Re = 20, representing 25% losses. Case 1 maintains more moderate performance, declining from 5.4 to 4.4 (18% loss). The first bend section reveals that Case 2 maintains dominance across the entire Reynolds number range, starting at Nu around 6.5 at Re = 5 and declining to 4.5 at Re = 20, representing 31% loss. Case 4 shows decline from 6.5 to 4.2 (35% loss), while Case 3 declines from 5.0 to 4.4 (12% loss) and Case 1 from 5.0 to 4.2 (16% loss). The 45° angle maintains more stable case ordering across Reynolds numbers compared to 30°. The second bend section shows Case 2 with substantial decline from 7.0 to 4.0 (43% loss), where it approaches baseline performance at high flow rates. Case 3 declines from 7.0 to 5.4 (23% loss), showing better retention. Case 4 drops from 7.5 to 4.7 (37% loss), while Case 1 shows decline from 5.5 to 4.7 (15% loss). At the steepest wire angle of 60° (**Figure 8**), Reynolds number effects reveal how dominant transverse magnetic field components respond to changing flow regimes. The channel region shows moderate Reynolds number dependencies. Case 2 displays negative trend, declining from 4.60 to 4.35 at Re = 20, representing 5% reduction. Cases 3 and 4 show variations around 4.40 to 4.45 across the Reynolds number range. Case 1 exhibits pattern with values between 4.33 and 4.38. The bend region demonstrates relatively modest Reynolds number dependence compared to shallower angles. Cases 3 and 4 both start around 7.5 at Re = 5 and while Case 4 declined substantially to approximately 4.5 at Re = 20, representing 40% losses, Case 3 shows decline from 7.5 to 5.0 (33% loss). Case 1 declines from 6.0 to 4.5 (25% loss). However, all of these trends are hardly monotonic as in the cases of other two angles, demonstrating the existence of competing flow regimes in flow mixing and consequently, heat transfer enhancement. The first bend section has monotonically decreasing trends for all the 4 current cases, with Case 4 staying above all for most of the Reynolds Number range. At Re = 5, Case 3 leads slightly with Nu around 7.3. Case 4 declines from 7.1 to 4.4 (39% loss), Case 2 from 6 to 4.3 (30% loss), Case 3 from 7.3 to 4.3 (41% loss), and Case 1 from 5.3 to 4.0 (27% loss) matching with the no-current case. By Re = 20, all cases have converged to Nu values between 4.0 and 4.5. The second bend section demonstrates distinctive. Case 2 starts at Nu around 9 at Re = 5 but experiences 56% decline to 3.8 at Re = 20, approaching baseline. Case 3 maintains steadier performance, declining from 7.2 to 6.2 (14% loss), showing exceptional Reynolds number resilience. Case 4 declines from 7.2 to 4.7 (34% loss), while Case 1 drops from 6.6 to 5.0 (24% loss). The dramatic Case 2 decline combined with Case 3 resilience establishes that wake region magnetic field enhancement at steep angles is highly configuration dependent and can be tuned not only for position in the channel, but also for flow rates, because monotonicity is broken because of local complex mixing.

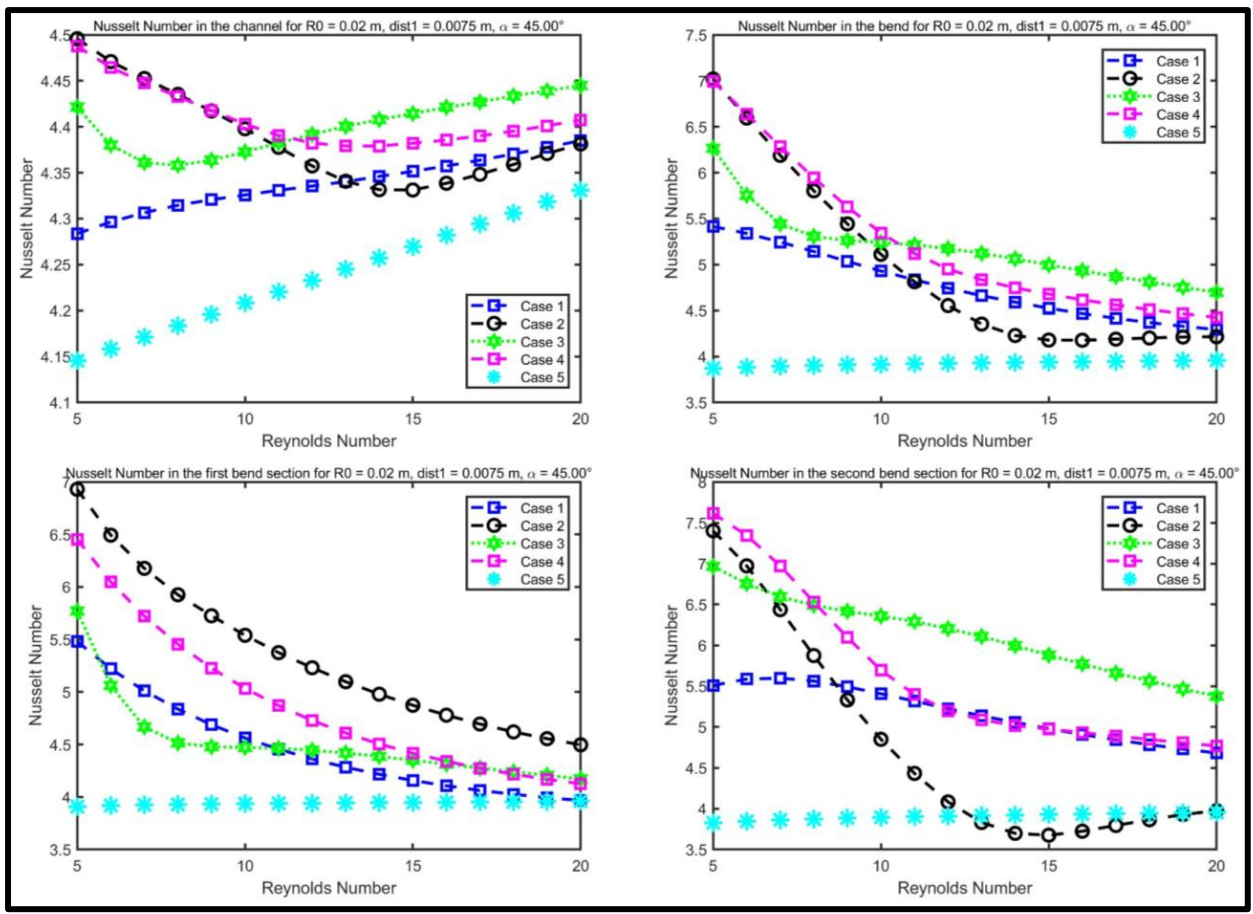

**Figure 7** Nusselt number variation with Reynolds number for $R_o = 0.02\ m$, $dist_1 = 0.0075\ m$, $\alpha = 45°$, $\phi = 0.05$.

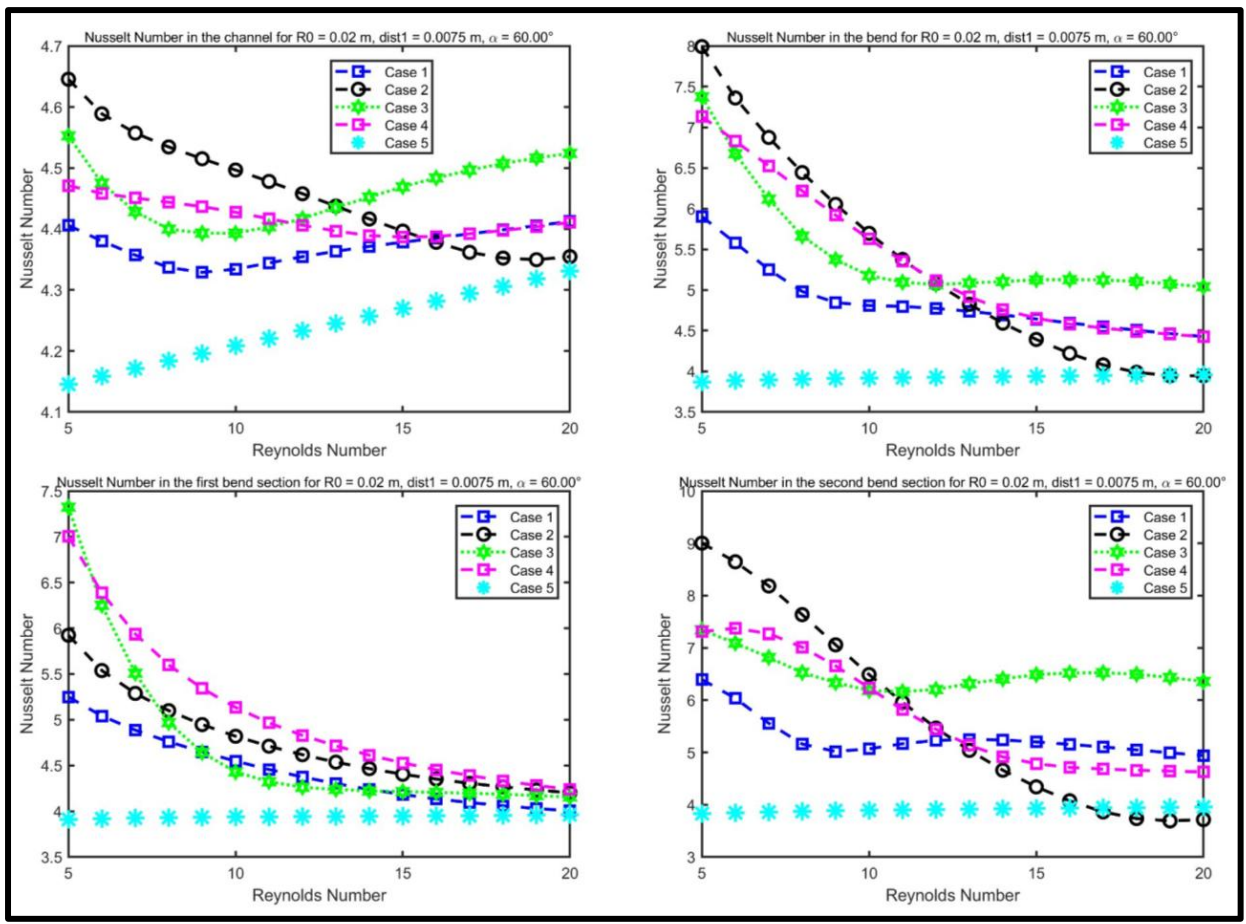

**Figure 8** Nusselt number variation with Reynolds number for $R_o = 0.02\ m$, $dist_1 = 0.0075\ m$, $\alpha = 60°$, $\phi = 0.05$.

**Figure 9** Shows the effect of Reynolds Number on all the 4 Nusselt numbers for the 5 Current cases for tight bend curvature ($R_o = 0.02\ m$), shallow wire angle ($\alpha = 30°$), tight wire positioning ($dist_1 = 0.0075\ m$) and higher nanoparticle concentration ($\phi = 0.1$). The Nusselt number sees a 30% increase for the whole channel for Re = 5, however, locally the Nusselt numbers see anywhere from 250% (for the whole bend) to 320% (for the first bend section) increase above the baseline, for case 3. Even at Re = 20, almost all cases end up significantly above the baseline, maintaining heat transfer enhancement by atleast 5% for the whole channel and 20% above the baseline locally (for the whole bend). Hence, the nano particle volume fraction is a very dominant factor in the heat transfer enhancement behavior of the ferrofluid flow. We will discuss this in mode detail in section 5.5.

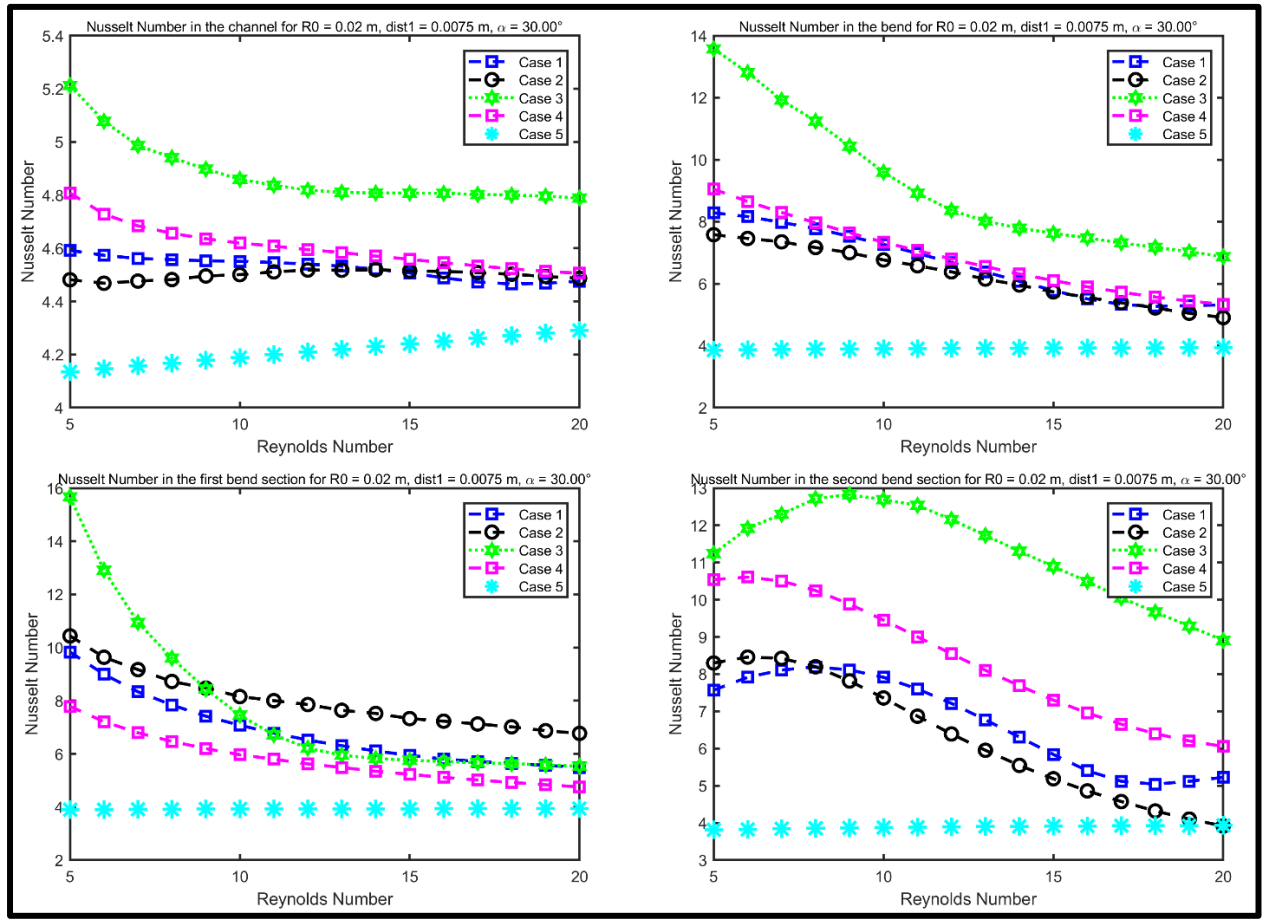

**Figure 9** Nusselt number variation with Reynolds number for $R_o = 0.02\ m$, $dist_1 = 0.0075\ m$, $\alpha = 30°$, $\phi = 0.1$

## 5.2 Effect of Bend Outer Radius on Heat Transfer Performance

Bend outer radius determines the curvature tightness of the flow path, which affects centrifugal forces and secondary flow patterns. This analysis examines three bend radii: $R_0 = 0.02\ m$ (tight curvature), 0.04 m (moderate curvature), and 0.06 m (gentle curvature), representing radius to channel diameter ratios of 2:1, 4:1, and 6:1 respectively. All parametric arrangements discussed here are for moderate Renolds Number (Re = 10), close wire positioning ($dist_1 = 0.0075\ m$), and 5% nanoparticle concentration. Three wire angles (30°, 45°, 60°) show how field orientation interacts with bend curvature effects. **Figure 10** shows the radius effect at shallow wire angle (30°) with close wire positioning. The whole channel displays moderate negative radius dependence for Case 3, declining from Nu around 4.64 at tight curvature to 4.48 at gentle curvature, representing approximately 3% reduction. Cases 1, 2, and 4 show minimal variation, remaining relatively flat around Nu = 4.35 to 4.40 across all radii. The baseline Case 5 maintains constant performance at Nu = 4.2. The bend region exhibits strong negative radius dependence across all Wire current cases. Case 3 shows the steepest decline, dropping from Nu around 7.1 at tight bends to 5.1 at gentle bends, representing 28% performance loss. Cases 1, 2, and 4 follow similar downward trends, each losing 20% to 25% as curvature becomes gentler. Case 2 declines from approximately

5.2 to 4.6, Case 4 from 5.5 to 4.7, and Case 1 from 5.8 to 4.8. The consistent negative slopes indicate that tight curvature is beneficial when wires are positioned close to the bend. The first bend section reveals dramatic radius sensitivity. Case 2 drops from Nu around 6.9 at $R_0 = 0.02\ m$ to approximately 4.8 at $R_0 = 0.06\ m$, a 30% reduction. Case 3 declines from 6.0 to 4.5, while Case 1 drops from 5.8 to 4.7. Case 4 exhibits more moderate decline from 4.9 to 4.5. The second bend section demonstrates continued strong radius dependence with Case 3 declining from Nu around 8.7 at tight curvature to 4.9 at gentle curvature, representing 44% performance loss.

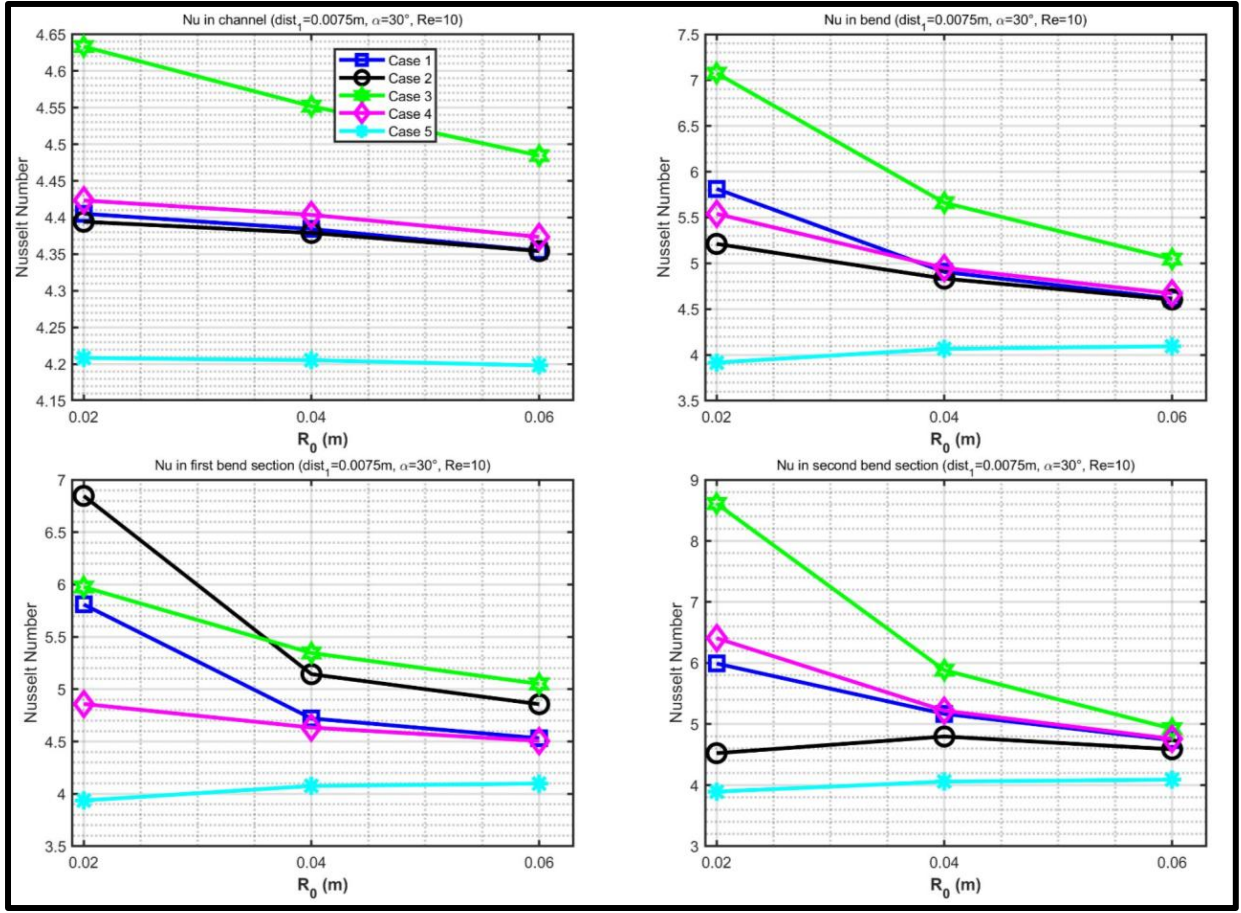

**Figure 10** Nusselt number variation with bend outer radius for $dist_1 = 0.0075\ m$, $\alpha = 30°$, $\phi = 0.05$, $Re = 10$.

Cases 1 and 4 show moderate declines around 25% to 30%, while Case 2 maintains relatively flat performance, which does come with minute non-monotonicity, and hence is a crucial observation. Regional Nusselt number variations with curvature are not always monotonic and can hence be leveraged for differential effects. However, the overwhelmingly consistent negative slopes across all regions at this shallow angle establish that tight bends are essential for maximizing magnetic field enhancement when wires are positioned close to the bend. At intermediate wire angle of 45° (**Figure 11**), radius effects show modified characteristics. The whole channel displays relatively flat behavior for most cases, with all ferrohydrodynamic configurations maintaining Nu values between 4.28 and 4.40 across the radius range. Case 3 shows slight non-monotonic trend from 4.37 to 4.38. The baseline remains constant around Nu = 4.2. The bend region demonstrates moderate negative radius dependence. Case 4 drops from Nu around 5.4 at $R_0 = 0.02\ m$ to 4.6 at $R_0 = 0.06\ m$, representing approximately 15% reduction. Cases 2 and 3 show similar declining trends of 12% to 14%, with Case 2 declining from 5.1 to 4.6 and Case 3 from 5.2 to 4.6. The slopes are less steep than at 30° angle. The first bend section exhibits interesting behavior where Case 2 maintains strongest performance at tight curvature (Nu around 5.6) but experiences steady decline to about 4.6 at gentle curvature, an 18% reduction.

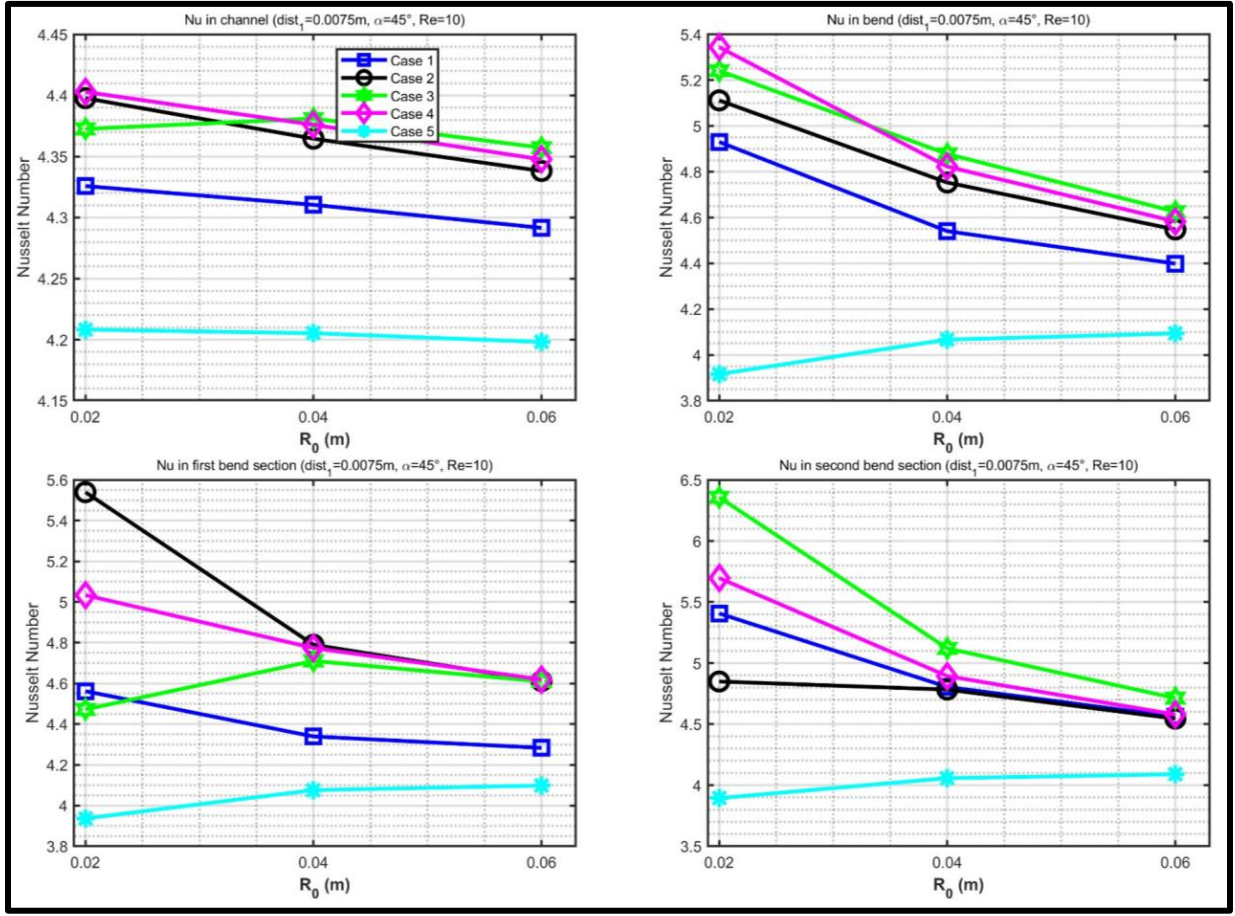

**Figure 11** Nusselt number variation with bend outer radius for $dist_1 = 0.0075\ m$, $\alpha = 45°$, $\phi = 0.05$, $Re = 10$.

Case 3 shows pattern with slight increase from $R_0 = 0.02\ m\ to\ 0.04\ m$ before declining at $R_0 = 0.06\ m$. Case 4 displays trend from 5.0 at tight curvature to 4.6 at gentle curvature. The second bend section shows Case 3 with negative radius dependence, declining from Nu around 6.4 at tight curvature to 4.7 at gentle curvature, representing 27% loss. Cases 1, 2, and 4 exhibit more moderate declines around 15% to 18%. The downward trends confirm that even at intermediate angles, tight bends provide better conditions for ferrohydrodynamic heat transfer enhancement. At the steepest wire angle (60°), **Figure 12** reveals different radius dependence patterns. The channel region shows diverse behaviors. Case 2 exhibits clear negative slope, declining from Nu around 4.50 at tight curvature to 4.37 at gentle curvature. Case 4 shows similar downward trend. Case 1 displays non monotonic behavior with a slight peak at moderate radius. Case 3 maintains relatively flat performance around Nu = 4.40. The varied response suggests that steep field orientations create complex interactions. The bend region demonstrates continued negative radius dependence but with reduced slopes compared to shallower angles. Cases 2 and 4 decline from Nu around 5.6 to 4.7, representing approximately 16% reduction. Case 3 shows modest decline from 5.2 to 4.7, about 10% loss. Case 1 declines from 4.8 to 4.6. The gentler slopes indicate that transverse magnetic field components are less sensitive to curvature changes. The first bend section reveals patterns where Case 4 shows strongest performance at tight curvature but gradual decline, maintaining competitive Nu values even at gentle curvature.

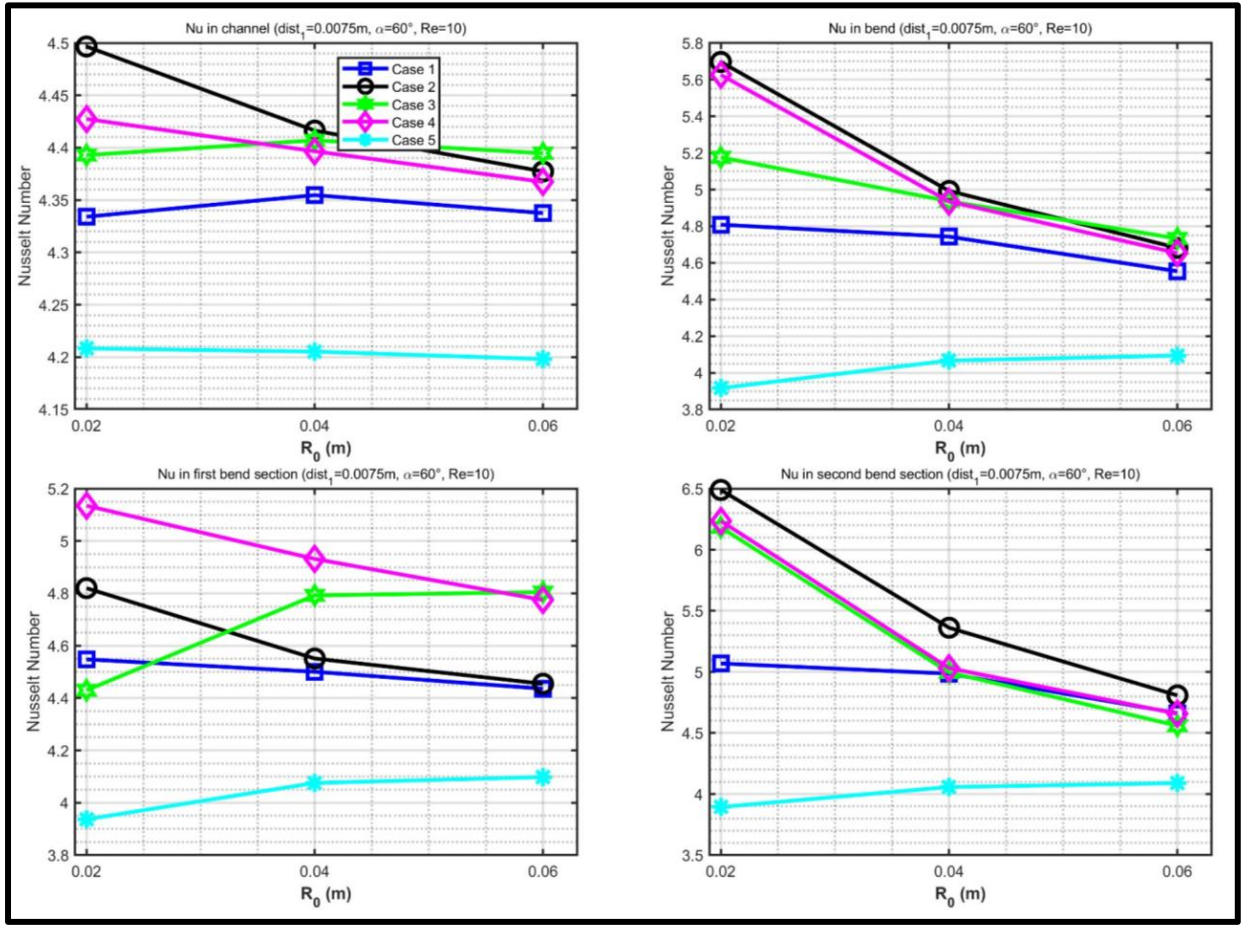

**Figure 12** Nusselt number variation with bend outer radius for $dist_1 = 0.0075\ m$, $\alpha = 60°$, $\phi = 0.05$, $Re = 10$.

Case 3 displays behavior with increase from $R_0 = 0.02\ m\ to\ 0.04\ m$ before declining at $R_0 = 0.06\ m$. Case 2 maintains relatively steady performance. The second bend section shows Case 2 with negative slope, declining from Nu around 6.5 at tight curvature to 4.8 at gentle curvature, a 26% reduction. Cases 3 and 4 show more moderate declines around 20% to 23%. The overall pattern at 60° suggests that steep wire angles create conditions where radius effects are more nuanced, with some cases showing reduced sensitivity compared to shallow angles.

## 5.3 Effect of Wire Angle on Heat Transfer Performance

Wire angle determines the orientation of magnetic fields relative to the flow direction, fundamentally affecting how magnetic body forces interact with the velocity field. This analysis examines seven wire angles from 30° to 60° in 5° increments, covering the full practical range from shallow to steep orientations. All tests use moderate Reynolds number (Re = 10), close wire positioning ($dist_1 = 0.0075\ m$), and 5% nanoparticle concentration. Three bend radii ($R_0 = 0.02, 0.04, 0.06\ m$) demonstrate how curvature interacts with field orientation effects. **Figure 13** shows angle effects at the tightest bend curvature with close wire positioning. The whole channel's Nusselt number displays diverse angle dependencies among cases. Case 3 exhibits strong negative slope, declining steadily from Nu around 4.64 at 30° to 4.40 at 60°, representing approximately 5% performance loss.

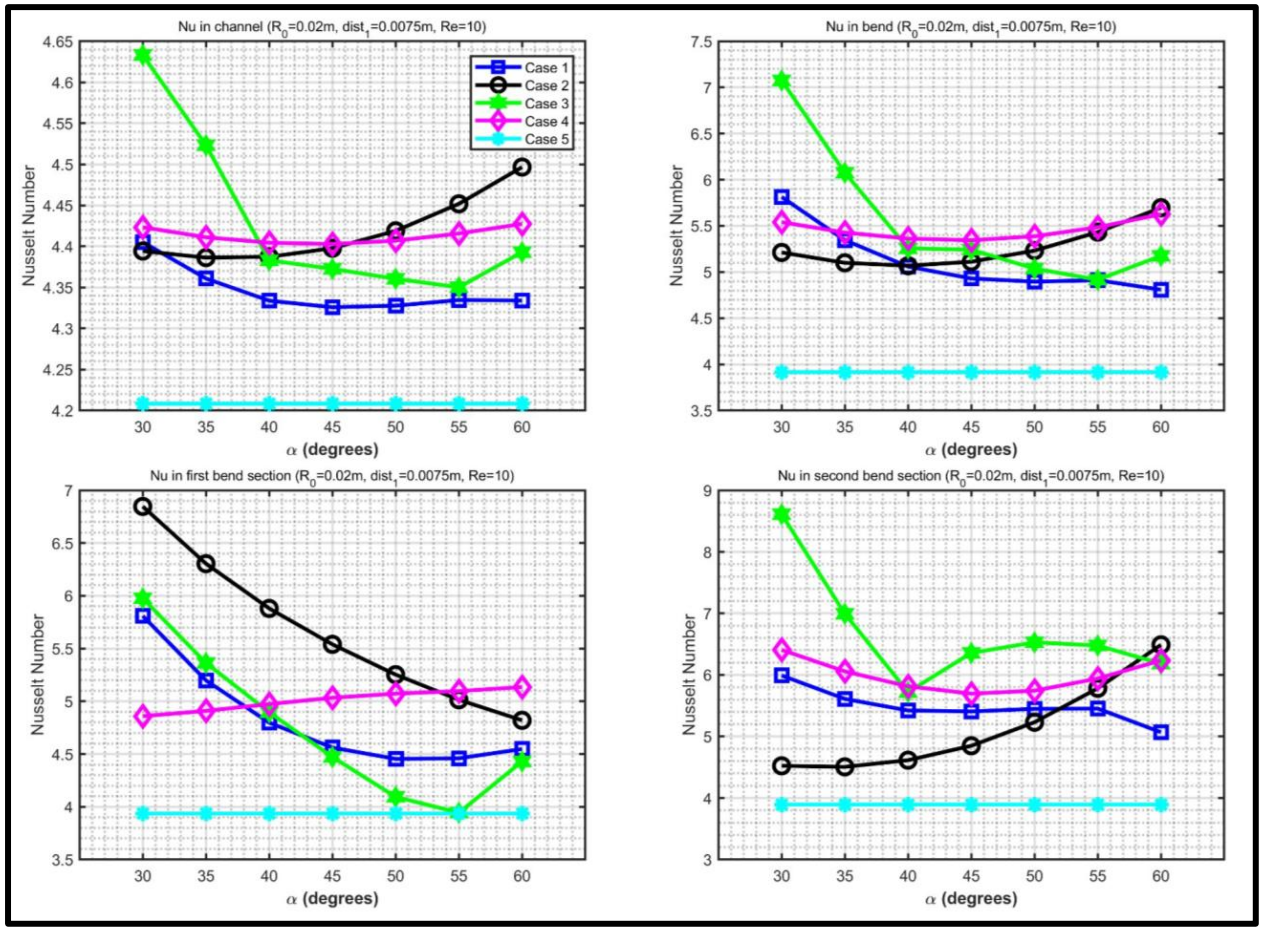

**Figure 13** Nusselt number variation with wire angle for $R_o = 0.02\ m$, $dist_1 = 0.0075\ m$, $Re = 10$, $\phi = 0.05$.

Case 2 shows opposite behavior with positive slope, increasing from 4.40 at 30° to 4.50 at 60°. Cases 1 and 4 maintain relatively flat profiles around Nu = 4.40 with minimal angle sensitivity. The baseline Case 5 remains constant at Nu = 4.2. The bend region reveals complex patterns. Case 3 displays pronounced V shaped behavior with peak performance at 30° (Nu around 7.1), declining to a minimum near 55° (Nu around 4.9), then recovering toward 60° (Nu around 5.3). Cases 2, and 4 show shallow parabolic patterns increasing slightly from 30° to 60°. The first bend section demonstrates dramatic angle effects. Case 2 shows steep negative slope, dropping from Nu around 6.9 at 30° to 4.8 at 60°, a 30% performance loss. Case 3 displays V shaped pattern with best performance at 30° and worst at intermediate angles around 50°. Case 4 maintains relatively steady performance. The second bend section shows Case 3 with clear V shaped dependence, achieving Nu around 8.7 at 30°, declining to minimum around 45° (Nu around 5.5), then partially recovering at 60° (Nu around 6.5). Case 2 exhibits non monotonic behavior with local minimum around 50°. These patterns establish that angle optimization is highly case specific, with Case 3 showing strong preference for angle extremes while Case 2 consistently favors shallow angles. At moderate bend curvature shown in **Figure 14**, angle effects show modified characteristics. The channel region displays smoother trends. Case 3 maintains negative slope but with reduced magnitude, declining from Nu around 4.55 at 30° to 4.38 at 55°, then recovering slightly at 60°. Case 1, 2 & 4 shows similar gentle parabolic slope (U-shaped profile) with lowest point occurring at 45°.

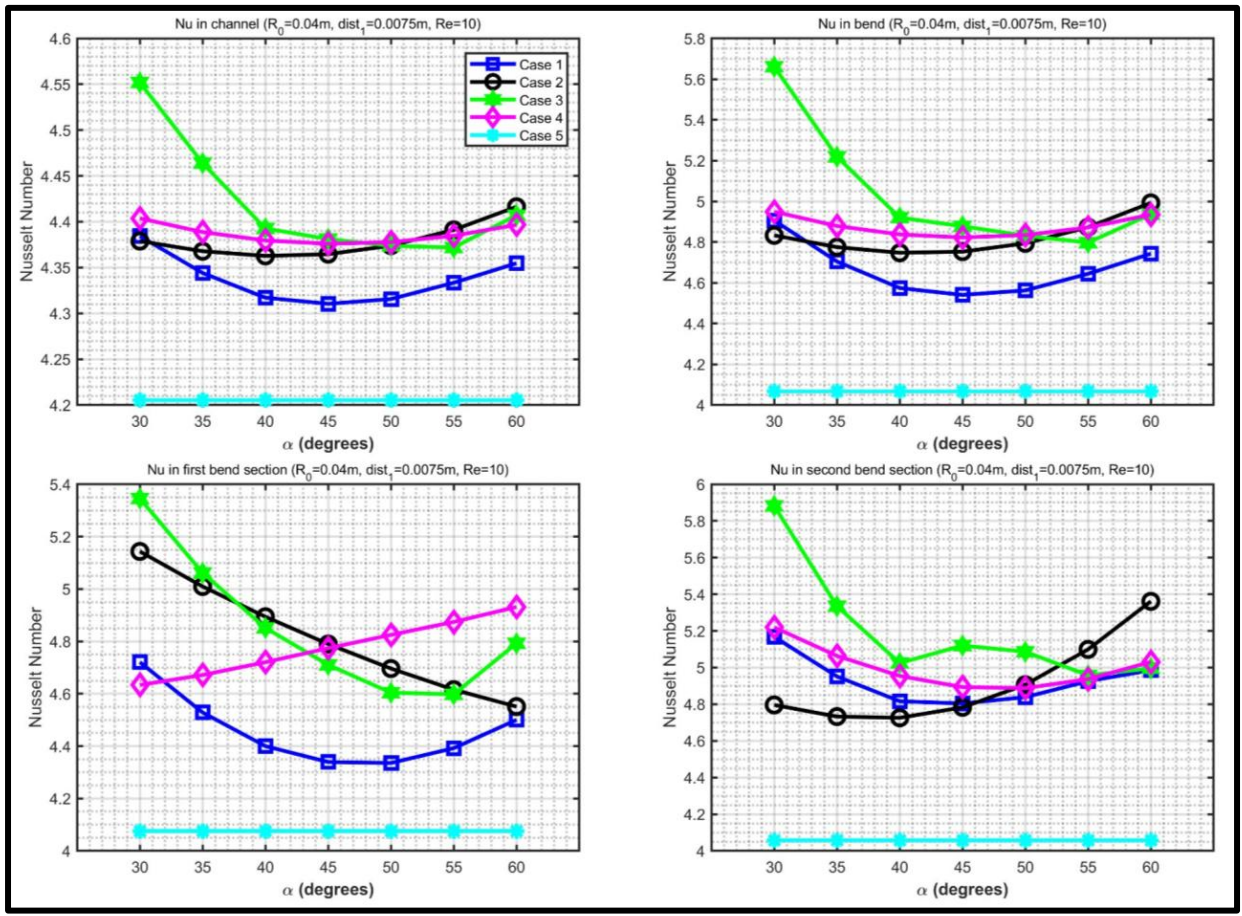

**Figure 14** Nusselt number variation with wire angle for $R_o = 0.04\ m$, $dist_1 = 0.0075\ m$, $Re = 10$, $\phi = 0.05$.

All cases converge more closely at steep angles compared to tight bend configuration. The bend region demonstrates continued V shaped behavior for Case 3, starting at Nu around 5.7 at 30°, declining to minimum near 55° (Nu around 4.7), then recovering to 4.9 at 60°. The V shape amplitude is reduced compared to tight bends. Cases 2 and 4 show relatively flat behavior. Case 1 displays modest declining trend from 4.9 to 4.7. The first bend section reveals crossovers among cases. Case 3 shows strong performance at shallow angles but experiences decline. Case 2 maintains more stable performance across angles. Case 4 exhibits rising trend from 4.6 at 30° to 4.9 at 60°. The second bend section shows Case 3 with persistent V shaped pattern, achieving Nu around 5.9 at 30°, minimum around 45° (Nu around 4.8), then recovery to 5.4 at 60°. Cases 1, 2, and 4 cluster more closely together. The reduced case differentiation at moderate curvature indicates that radius and angle effects are coupled. At the gentlest bend curvature (**Figure 15**), angle effects become most attenuated. The channel region shows compressed variations with all Current cases maintaining Nu values between 4.30 and 4.50 across the entire angle range. Case 3 still shows slight negative trend from 4.49 at 30° to 4.38 at 60°, but the total variation is only 2.5%. Cases 1, 2, and 4 display nearly flat or weakly U-shaped patterns.

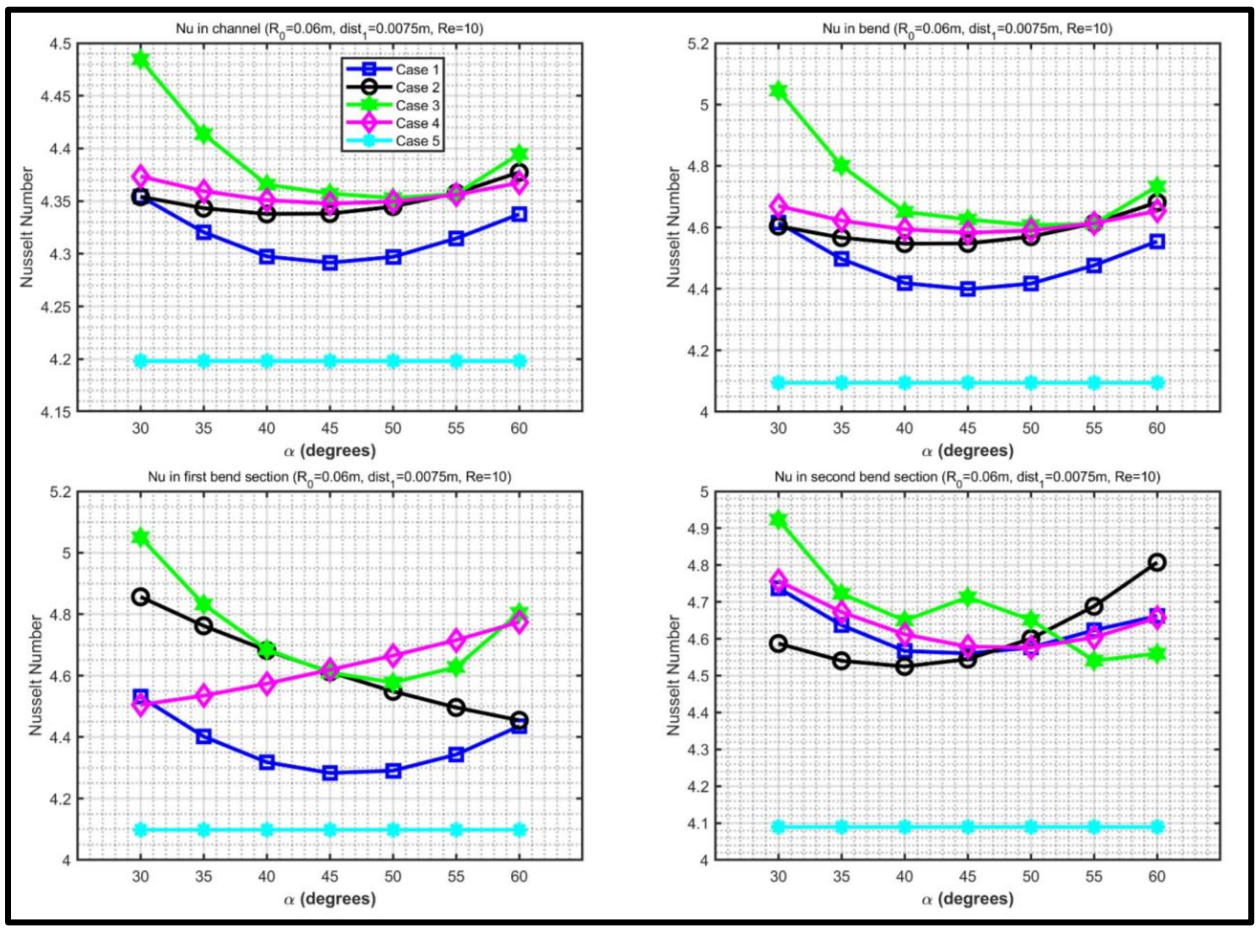

**Figure 15** Nusselt number variation with wire angle for $R_o = 0.06\ m$, $dist_1 = 0.0075\ m$, $Re = 10$, $\phi = 0.05$.

The minimal angle sensitivity at gentle curvature suggests flow smoothing takes over local mixing effects. The bend region demonstrates continued evidence of V shaped behavior for Case 3, though with further reduced amplitude. Starting at Nu around 5.1 at 30°, Case 3 declines to minimum near 45° (Nu around 4.6), then recovers to 4.7 at 60°. Other cases show relatively flat behavior. The first bend section reveals patterns where Case 3 maintains modest performance advantage at shallow angles (Nu around 5.1 at 30°) but shows clear minimum around 45° before partial recovery. Case 2 exhibits declining trend from 4.8 to 4.5. Case 4 shows rising trend, performing better at steep angles with Nu increasing from 4.5 at 30° to 4.8 at 60°. The second bend section shows Case 3 with weakest V shape amplitude among all three radii, varying only between Nu = 4.5 and 4.9 across the angle range. Case 2 displays gradual rise from 4.6 at 30° to 4.8 at 60°. All cases converge closely at steep angles, indicating that when both curvature and field orientation reach their extremes, performance differences between magnetic field configurations diminish substantially.

## 5.4 Effect of Wire Distance on Heat Transfer Performance

Wire distance from the bend determines the magnetic field strength at the location where heat transfer occurs, following an inverse square relationship where field intensity decreases rapidly with distance. This analysis examines two wire positions: $dist_1 = 0.0075\ m$ (close positioning) and $dist_1 = 0.0100\ m$ (distant positioning), representing a 33% increase in separation distance. All tests use tight bend curvature ($R_0 = 0.02$ m), moderate Reynolds number (Re = 10), and 5% nanoparticle concentration. Three wire angles (30°, 45°, 60°) demonstrate how field orientation interacts with distance effects. **Figure 16** shows distance effects at shallow wire angle with tight bend curvature. The channel region displays universal negative distance dependence across all current cases. Case 3 shows the steepest decline, dropping from Nu around 4.64 at close positioning to 4.37 at distant positioning, representing approximately 6% performance loss. Cases 1, 2, and 4 show similar declining trends with losses between 3% and 5%. The baseline Case 5 remains constant at Nu = 4.2. The bend region demonstrates severe distance sensitivity. Case 3 shows catastrophic decline from Nu around 7.1 at close positioning to 5.3 at distant positioning, representing 25% performance loss. Cases 1, 2, and 4 experience similar severe losses. All cases converge toward Nu values around 4.2 to 4.3 at distant positioning. The first bend section reveals the most severe distance penalties. Case 2 experiences catastrophic decline from Nu around 6.9 at close positioning to 5.3 at distant positioning, representing 23% performance loss. Case 3 shows dramatic decline from 6.0 to 5.1 (15% loss), Case 1 from 5.8 to 4.2 (28% loss), and Case 4 from 4.9 to 3.6 (27% loss). The second bend section demonstrates continued severe distance dependence with Case 3 declining from Nu around 8.7 at close positioning to 5.7 at distant positioning, representing 34% performance loss. The universal steep negative slopes establish distance as the single most critical parameter.

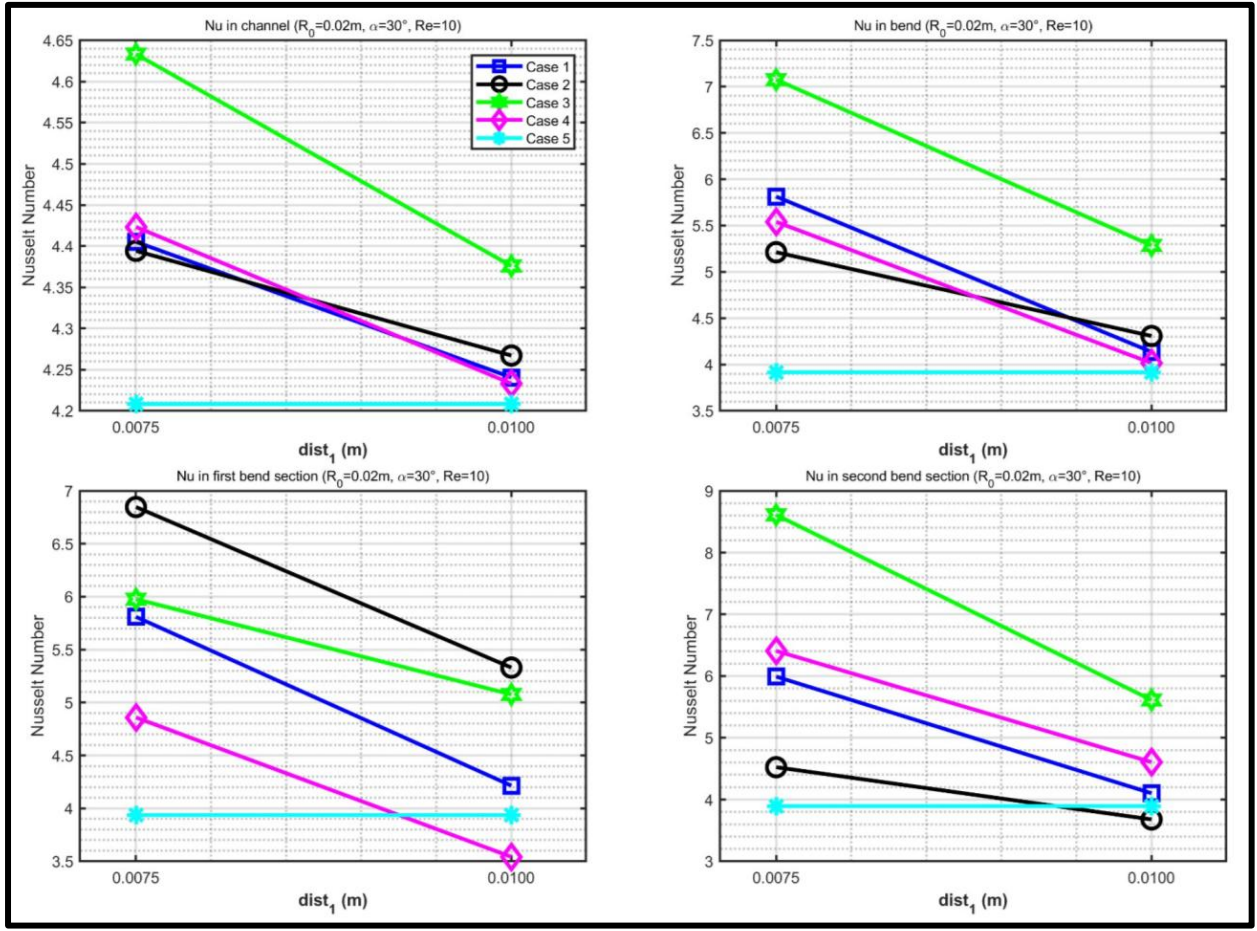

**Figure 16** Nusselt number variation with wire distance for $R_o = 0.02\ m$, $\alpha = 30°$, $Re = 10$, $\phi = 0.05$

At intermediate wire angle of 45° (**Figure 17**), distance effects remain severe. The channel region displays universal negative distance dependence with parallel slopes. Case 2 and Case 4 show nearly identical behavior, both declining from Nu around 4.40 at close positioning to 4.22 at distant positioning, representing 4% loss. Case 3 follows similar trend, dropping from 4.37 to 4.23. All cases converge very closely at distant positioning. The bend region demonstrates severe distance sensitivity. Case 4 shows steepest decline from Nu around 5.4 at close positioning to 4.0 at distant positioning, representing 26% performance loss. Cases 2 and 3 show similar catastrophic declines. All cases converge extremely closely at distant positioning. The first bend section reveals varied distance effects. Case 2 maintains strongest performance at close positioning (Nu around 5.5) but experiences steep 22% decline to 4.3 at distant positioning. Case 4 shows dramatic decline from 5.1 to 3.5, representing 31% loss. Case 3 declines from 4.5 to 4.0 (11% loss). The second bend section shows Case 3 with strong negative slope, declining from Nu around 6.4 at close positioning to 4.5 at distant positioning, representing 30% loss. Case 2 experiences 26% decline, while Case 4 shows 22% loss.

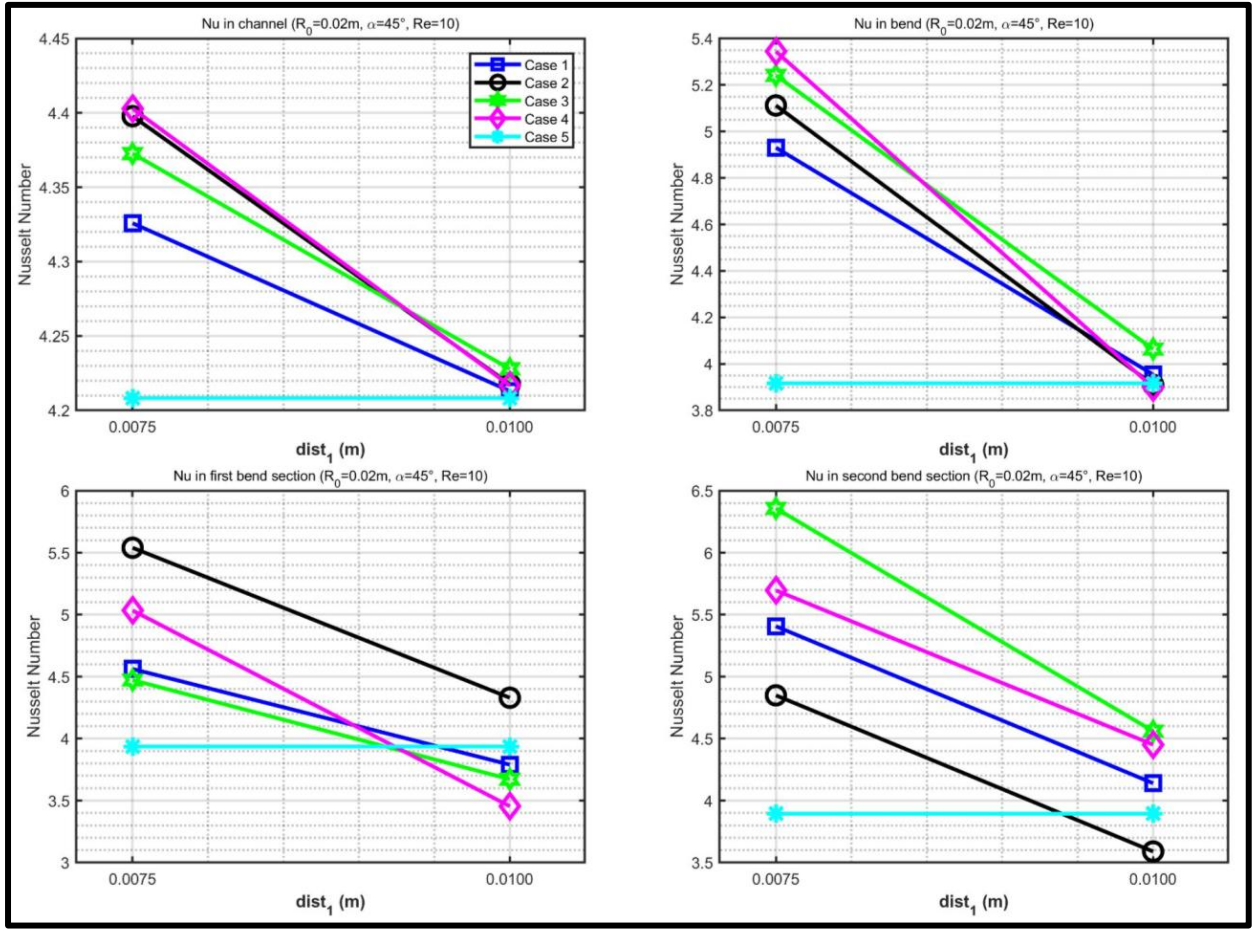

**Figure 17** Nusselt number variation with wire distance for $R_o = 0.02\ m$, $\alpha = 45°$, $Re = 10$, $\phi = 0.05$

At the steepest wire angle (60°), **Figure 18** reveals that distance effects remain severe. The channel region shows universal negative distance dependence. Case 2 exhibits the steepest decline, dropping from Nu around 4.50 at close positioning to 4.25 at distant positioning, representing 6% performance loss. Cases 3 and 4 show similar behavior. All cases converge closely at distant positioning. The bend region demonstrates catastrophic distance sensitivity. Case 4 shows dramatic decline from Nu around 5.7 at close to 3.9 at distant positioning, representing 32% performance loss. Case 2 experiences similar 31% decline. The exceptionally steep slopes indicate that transverse magnetic field components are critically dependent on field strength. The first bend section reveals dramatic distance penalties. Case 4 experiences catastrophic decline from Nu around 5.2 at close positioning to 3.5 at distant positioning, representing 33% performance loss. Case 2 shows 17% decline. The second bend section shows Case 2 with the most severe distance dependence observed in any configuration, declining from Nu around 6.5 at close positioning to 3.7 at distant positioning, representing catastrophic 43% performance loss. Case 4 shows 28% decline, Case 3 experiences 24% loss. The universal convergence toward low Nu values establishes that steep angles absolutely require close wire positioning to maintain any performance advantage.

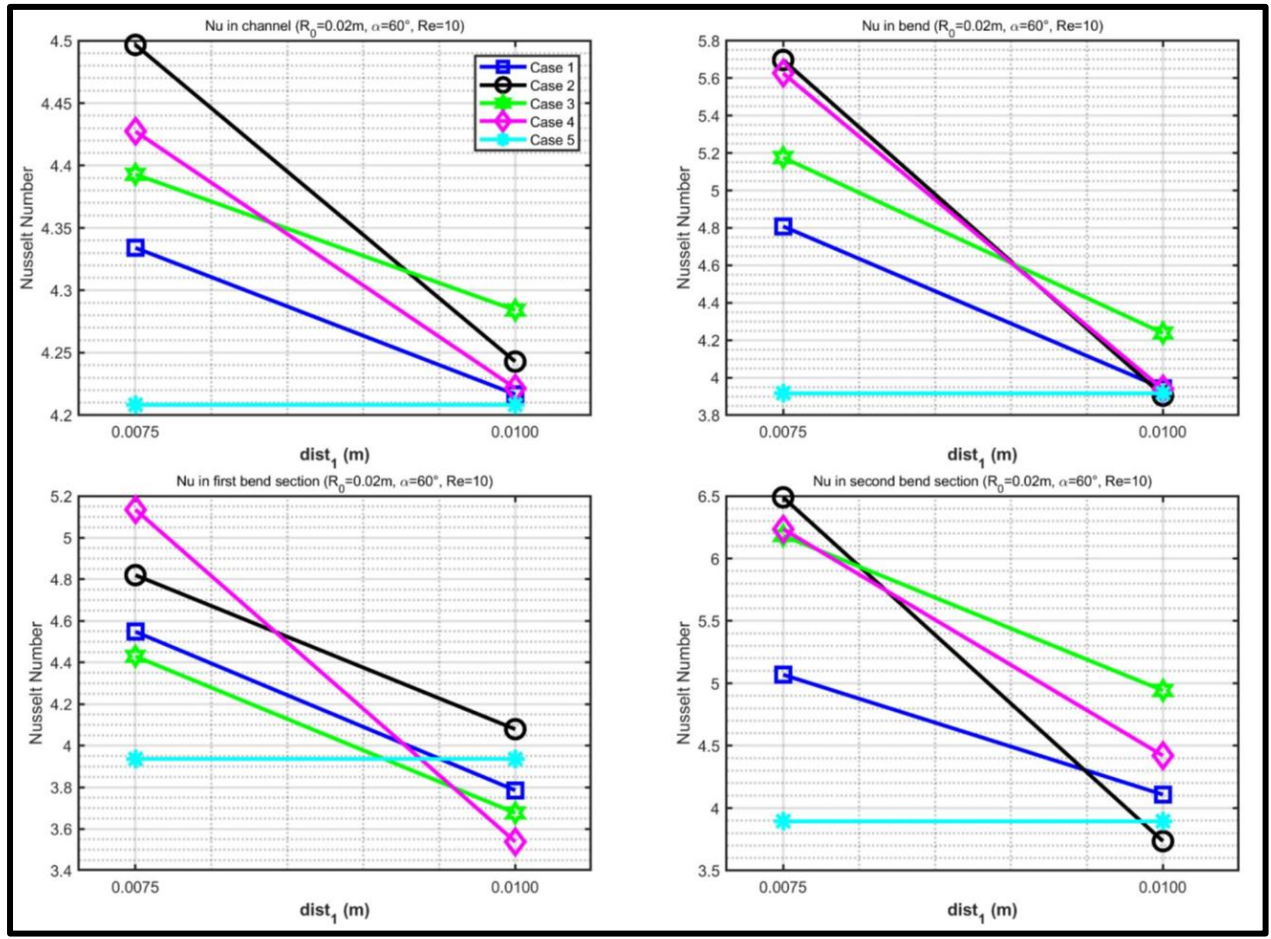

**Figure 18** Nusselt number variation with wire distance for $R_o = 0.02\ m, \alpha = 60°, Re = 10,\ \phi\ =\ 0.05$

## 5.5 Effect of Nanoparticle Volume Fraction on Heat Transfer Performance

Nanoparticle volume fraction determines both the thermal conductivity of the ferrofluid and its magnetic susceptibility, creating dual enhancement mechanisms. This analysis examines two concentrations: $\phi\ =\ 0.05$ (5% by volume) and $\phi\ =\ 0.10$ (10% by volume), representing a doubling of nanoparticle loading. Doubling the concentration increases thermal conductivity by approximately 10% to 15% but more importantly increases magnetic susceptibility by nearly 133%, thereby non-linearly increasing the Kelvin body forces that drive magnetic field induced mixing. However, this does come with a penalty in the heat capacity of the ferrofluid, which reduces by 15%, by doubling the nanoparticle concentration. All cases presented here use tight bend curvature (R₀ = 0.02 m), close wire positioning ($dist_1\ =\ 0.0075\ m$), and moderate Reynolds number (Re = 10). Three wire angles (30°, 45°, 60°) demonstrate how field orientation interacts with concentration effects.

**Figure 19** shows concentration effects at shallow wire angle with optimal geometric conditions. The channel region displays universal positive concentration dependence. Case 3 shows the strongest concentration sensitivity, increasing from Nu around 4.64 at $\phi\ =\ 0.05$ to 4.87 $\phi\ =\ 0.1$

, representing 5% enhancement. Case 1 increases from 4.40 to 4.55, Case 2 from 4.40 to 4.50, and Case 4 from 4.43 to 4.62. The baseline Case 5 shows minimal variation. The bend region demonstrates dramatically amplified concentration effects. Case 3 exhibits exceptional concentration sensitivity, surging from Nu around 7.05 at $\phi = 0.05$ to 9.70 at $\phi = 0.1$, representing remarkable 38% enhancement. Case 2 shows 30% improvement, Case 4 increases 30%, and Case 1 gains 25%. All ferrohydrodynamic cases show steeper slopes compared to the channel. The first bend section reveals case specific concentration sensitivities. Case 2 maintains dominance but shows moderate 19% enhancement. Case 3 displays 7% improvement. Case 1 shows 21% gain, while Case 4 increases 22%. The second bend section demonstrates the maximum concentration effects. Case 3 achieves extraordinary enhancement, soaring from Nu around 8.70 at φ = 0.05 to 12.80 at φ = 0.10, representing phenomenal 47% improvement. Case 4 shows 43% enhancement, Case 1 gains 33%, while Case 2 increases 64%. These dramatic improvements confirm that magnetic field effects dominate concentration benefits.

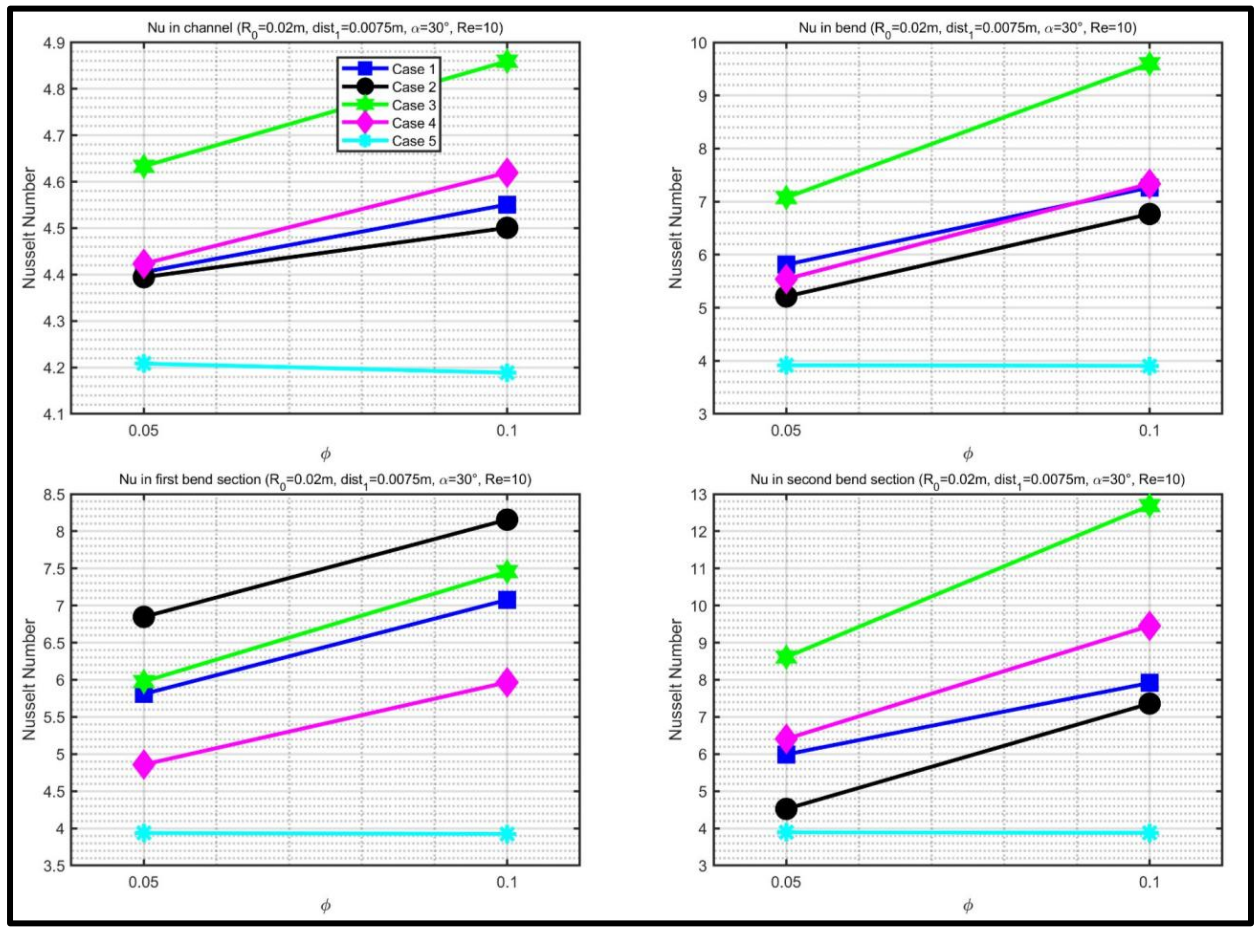

**Figure 19** Nusselt number variation with nanoparticle volume fraction for $R_o = 0.02\ m$, $\alpha = 30°$, $Re = 10$, $dist_1 = 0.0075$

At intermediate wire angle of 45° (**Figure 20**), concentration effects show modified characteristics with more uniform responses. The channel region displays convergence of concentration sensitivities. Cases 2, 3, and 4 show nearly identical behavior, all increasing to approximately 4.50 to 4.58 at φ = 0.10, representing uniform 3% to 4% enhancements. Case 1 shows 2.7% improvement. The tight clustering indicates that 45° orientation creates standardized channel performance. The bend region reveals case reordering with concentration changes. At φ = 0.05, Cases 2 and 3 perform similarly, but at φ = 0.10, Case 4 emerges dominant achieving Nu around 7.35, representing 36% enhancement. Case 2 shows 35% improvement, while Case 3 displays 25% gain. Case 1 increases 20%. This crossover indicates that single wire configurations benefit from increased nanoparticle loading at intermediate angles. The first bend section displays uniform concentration effects. Case 2 maintains leadership, increasing 22%. Case 3 shows 19% enhancement, Case 4 gains 28%, and Case 1 increases 22%.

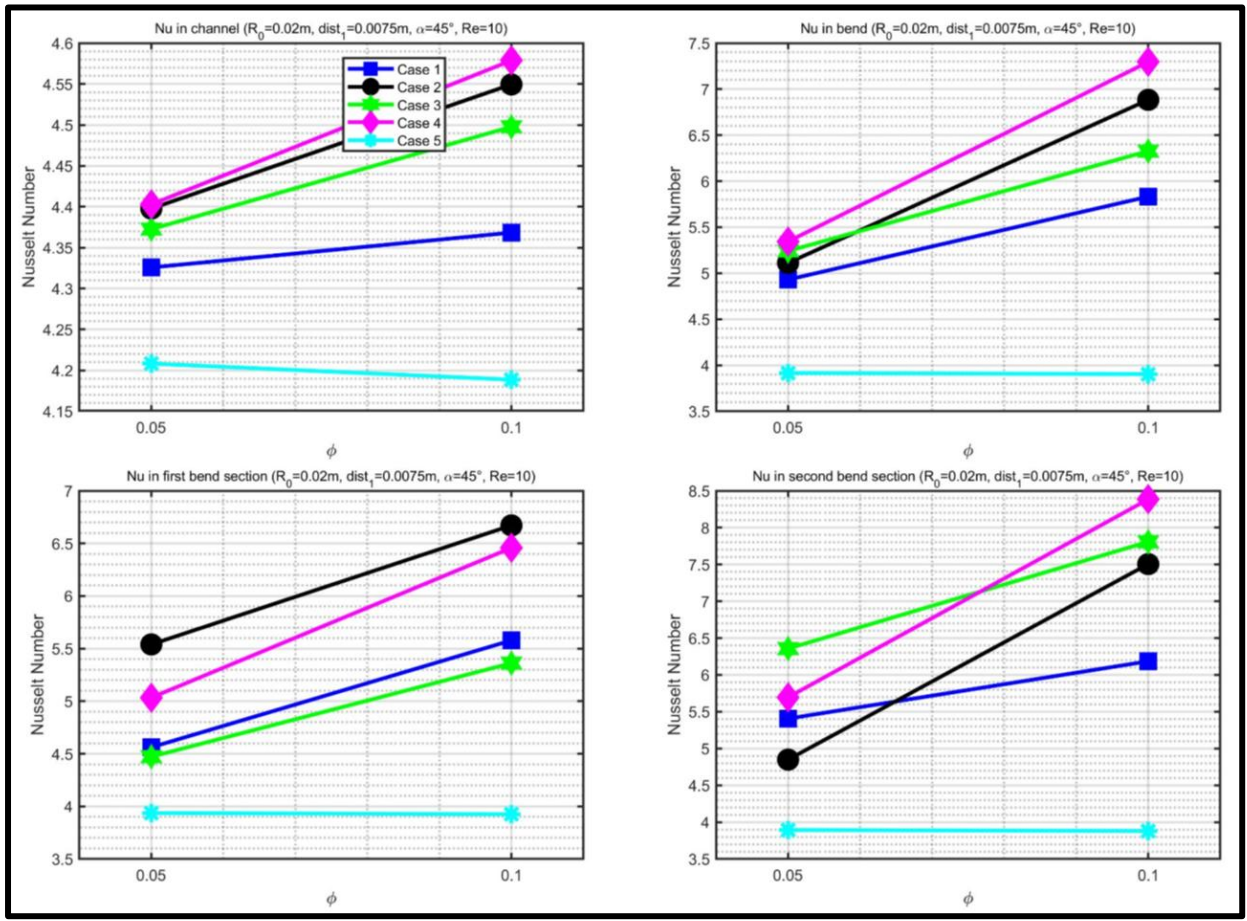

**Figure 20** Nusselt number variation with nanoparticle volume fraction for $R_o = 0.02\ m, \alpha = 45°, Re = 10, dist_1 = 0.0075$

The second bend section shows Case 4 exhibiting extraordinary concentration sensitivity, surging from Nu around 5.70 at φ = 0.05 to 8.40 at φ = 0.10, representing phenomenal 47% enhancement. Case 3 shows 23% gain, Case 2 increases 53%, and Case 1 gains 13%.

At the steepest wire angle (60°), **Figure 21** reveals concentration effects under dominant transverse field conditions. The channel region shows the weakest concentration sensitivities. Case 2 exhibits the strongest response with only 3.1% enhancement, while Cases 3 and 4 show 2.4% to 2.7% improvements, and Case 1 displays 1.4% gain. The reduced concentration effects indicate that steep wire orientations create field distributions less conducive to concentration dependent enhancement in straight sections. The bend region demonstrates moderate but remarkably uniform concentration dependencies. Case 4 achieves 36% enhancement, followed by Case 2 with 37% gain, and Case 3 showing 35% improvement. Case 1 displays 12% gain. The near parity in concentration response across Cases 2, 3, and 4 reflects the homogenizing effect of steep angles. The first bend section reveals differential concentration sensitivities. Case 4 dominates the concentration response, increasing dramatically, representing 35% enhancement. Case 3 shows 26% gain, while Case 2 exhibits 15% improvement. Case 1 increases 16%. The second bend section demonstrates substantial concentration effects. Case 2 shows the strongest response, increasing from Nu around 6.50 to 9.15, representing 41% enhancement.

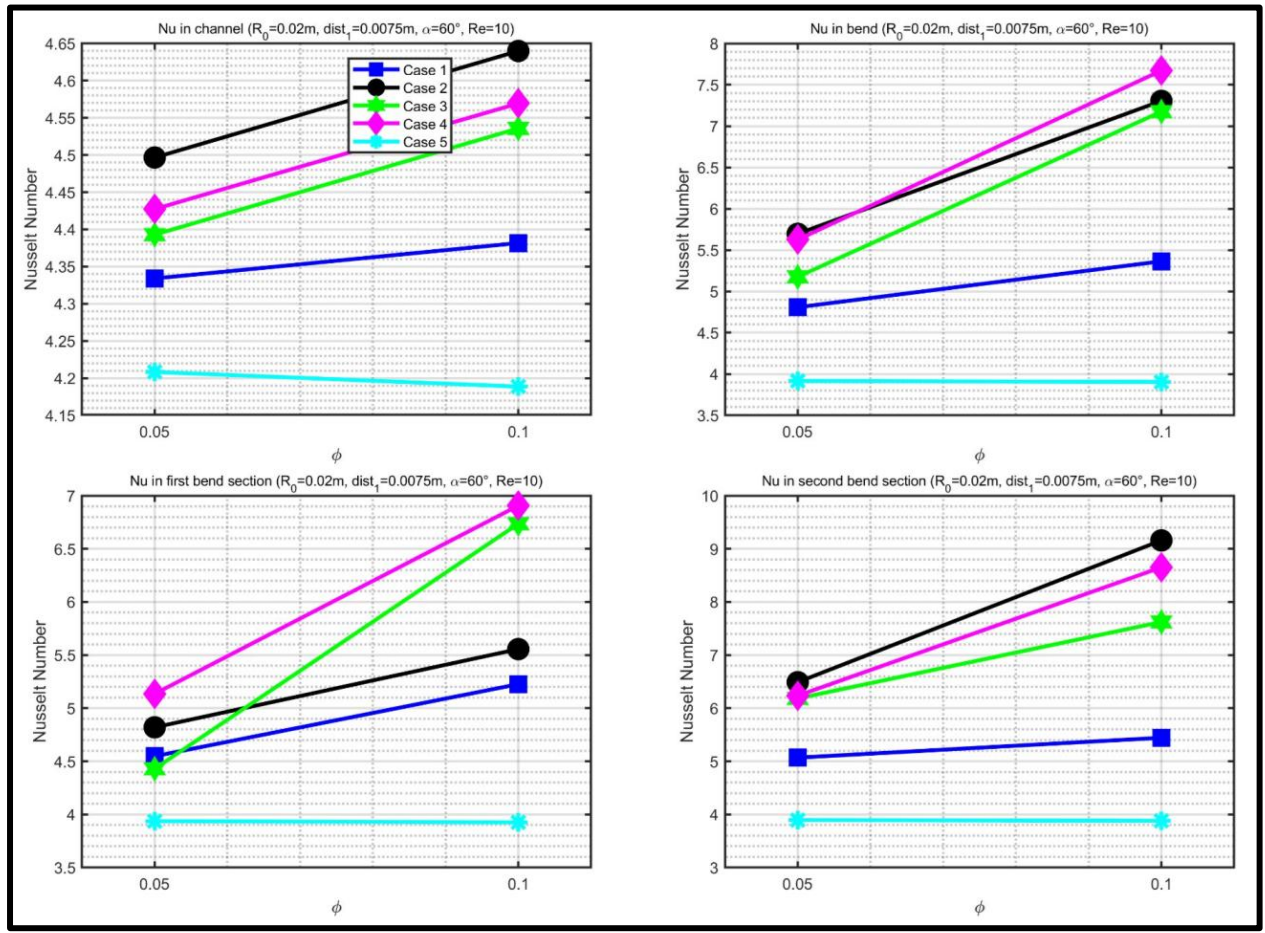

**Figure 21** Nusselt number variation with nanoparticle volume fraction for $R_o = 0.02\ m, \alpha = 60°, Re = 10, dist_1 = 0.0075$

Case 4 gains 42%, while Case 3 increases 43%. Case 1 shows 8% improvement. The similar enhancement magnitudes across Cases 2, 3, and 4 reinforce the homogenization pattern at steep angles.

# 6. CONCLUSIONS

This study examined ferrohydrodynamic heat transfer enhancement in curved channels through systematic parametric analysis of five key variables: Reynolds number, bend radius, wire angle, wire distance, and nanoparticle volume fraction. The investigation covered all the unique configurations using two wire distances, three bend radii, seven wire angles, four Reynolds numbers, two nanoparticle concentrations, and five wire current cases (excluding the ones which are symmetric). Results show that proper parameter selection can achieve heat transfer enhancements up to 30% globally (for the whole channel) and up to 300% (locally for the bend) compared to baseline flow without magnetic field effects. The main findings are summarized below:

- Reynolds number strongly affects performance with optimal operation at Re = 5. Low Reynolds numbers (Re = 5) allow magnetic body forces to dominate, while high Reynolds numbers (Re = 20) cause 35% to 58% performance loss compared to Re = 5 depending on configuration. Case 2 at 60° wire angle showed catastrophic 58% decline in the second bend section as inertia overwhelmed magnetic effects.

- Bend radius effects show strong coupling with wire angle. Tight bends ($R_o$ = 0.02 m) perform best when wires are close, with performance dropping 10% to 44% at gentle curvature ($R_o$ = 0.06 m). Case 3 experienced the largest 44% loss in the second bend section at 30° angle.

- Wire angle optimization reveals complex behavior. Case 3 with opposing currents exhibits V shaped performance with peaks at 30° or 60° and valleys at 40° to 55°. Case 2 consistently favors shallow angles below 35°. The performance valley at intermediate angles represents 20% to 28% reduction compared to optimal angles.

- Wire distance proved to be the most critical parameter. All cases showed universal negative distance dependence with 2% to 43% performance loss when distance increased 33%. Case 2 at 60° angle experienced catastrophic 43% decline in the second bend section. Magnetic field strength follows inverse square law, making close positioning essential.

- Nanoparticle concentration showed universal positive effects. Doubling concentration from 5% to 10% increased heat transfer by 2% to 64% depending on configuration. Case 2 achieved maximum 64% enhancement in the second bend section at 30° angle. The improvement comes primarily from doubled magnetic susceptibility rather than modest 10% to 15% thermal conductivity gains.

- The second bend section showed strongest response to nanoparticle concentration, with enhancements between 7% and 64%. Case 3 at steep angles maintained best Reynolds number resilience with only 7% decline compared to 58% for other cases.

- The best overall configuration used opposing wire currents (Case 3), tight bend radius ($R_0$ = 0.02 m), close wire spacing ($dist_1 = 0.0075\ m$), shallow angle (30°), moderate flow (Re = 5), and high nanoparticle loading ($\phi = 0.10$). This setup achieved Nu around 16 in the first bend section, representing approximately 320% enhancement over baseline.

The results reveal that parameters interact in complex ways that make independent optimization ineffective. Several coupling rules emerged that must be followed for good performance. Wire distance must be minimized as the priority since it affects all cases universally and severely. Wire angle must avoid the 40° to 50° range where Case 3 shows substantial performance degradation. Nanoparticle concentration should be higher (as permitted by dilute ferrofluid property relationships) since it always helps performance through enhanced magnetic susceptibility. Among all parameters, wire distance matters most because of the strong inverse square relationship between distance and magnetic field strength.

The physical mechanisms behind these trends are well understood. Kelvin body forces that drive the enhancement depend on both magnetic field gradients and field strength squared, so they decay rapidly with distance from the wires. At low Reynolds numbers, these magnetic forces control the flow behavior. As Reynolds number increases beyond 15, inertial forces become dominant and ferrohydrodynamic effects weaken substantially. The first bend section shows strong enhancement because thermal boundary layers are thin and easily disrupted by magnetic body forces. The second bend section responds most strongly to nanoparticle concentration because recirculation patterns in this region amplify the effects of stronger magnetic response. When concentration doubles, magnetic susceptibility doubles while thermal conductivity improves only 10% to 15%, yet the combined effect can reach 64% enhancement because the mechanisms work synergistically.

Apart from these general observations, it is highly intriguing that the interplay between the parameters change the evolution of Nusselt numbers (and by extension the heat transfer enhancement) non-monotonically, often defying existing patterns. This can be used for targeted cooling applications, where granular tunability to mixing and heat transfer enhancement would help achieve the requisite cooling performance by a simple tuning of parameters. Finally, this work can be extended further by exploring channels with other than 90° bends, using different magnetic objects to induce fields in the flow, exploring different nanoparticles, magnetocaloric effects etc.

# Nomenclature

| | | | |
|---|---|---|---|
| $w_0$ | Width of the channel | $\alpha$ | Angle made by line joining the wires with horizontal |
| $L_0$ | Length of the channel | $R_o$ | Outer radius of the bend |
| $\chi_m$ | Magnetic Susceptibility of the ferrofluid | $T$ | Temperature |
| $U$ | Inlet Velocity | $Nu_{ch}$ | Nusselt Number for the whole channel |
| $\vec{V}$ | Velocity vector | $Nu_{bend}$ | Nusselt number for the bend |
| $Re$ | Flow Reynolds Number | $Nu_{bend1}$ | Nusselt number in the first bend section |
| $u$ | x component of velocity | $Nu_{bend2}$ | Nusselt number in the second bend section |
| $v$ | y component of velocity | $k_p$ | Thermal conductivity of particles |
| $\eta$ | Dynamic viscosity of the ferrofluid | $k_L$ | Thermal conductivity of carrier liquid |
| $\rho$ | Density of the ferrofluid | $\eta_L$ | Dynamic viscosity of carrier liquid |
| $I_1 \& I_2$ | Current magnitude in the current carrying wires | $\rho_L$ | Density of the carrier liquid |
| $M_s$ | Initial Magnetization of the ferrofluid | $\rho_p$ | Density of ferromagnetic particles |
| $dist_1$ | Distance of the wire center from the channel center | $C_{pL}$ | Specific heat of the carrier liquid |
| $N_{cell}$ | Mesh refinement parameter | $C_{pp}$ | Specific heat of ferromagnetic particles |
| $H$ | Magnetic field generated from the wires | $d$ | Diameter of ferromagnetic particles |
| $B$ | Magnetic Flux Density | $\mu_B$ | Bohr Magneton ($9.27 \times 10^{-24} Am^2$) |
| $p$ | Pressure | $T_{ref}$ | Reference temperature |
| $\mu_0$ | Permeability of free space $4\pi \times 10^{-7} \, N/A^2$ | $\phi$ | Volume fraction of ferrofluid |
| $m$ | Magnetic moment of ferromagnetic particles | $k_B$ | Boltzmann constant ($1.38 \times 10^{-23} J/K$) |
| $n$ | Number density of ferromagnetic particles | $M_d$ | Saturation magnetization for magnetite (423 kA/m) |